\documentstyle[12pt,epsf]{article}
\newcommand\VV{\setbox0=\hbox{V}\hbox{\rm V\raise\ht0
  \hbox to0pt{\hss\vbox to0pt{\hbox{v}\vss}}}}
\def\slashchar#1{\setbox0=\hbox{$#1$}           
   \dimen0=\wd0                                 
   \setbox1=\hbox{/} \dimen1=\wd1               
   \ifdim\dimen0>\dimen1                        
      \rlap{\hbox to \dimen0{\hfil/\hfil}}      
      #1                                        
   \else                                        
      \rlap{\hbox to \dimen1{\hfil$#1$\hfil}}   
      /                                         
   \fi}                                         %
\thicklines
\begin{document}
\hfill hep-ph/0002127
\begin{center}
{\large\bf  $B_c$ Decays and Lifetime in QCD Sum Rules}\\
\vspace*{4mm}
{V.V.Kiselev\footnote{E-mail: kiselev@th1.ihep.su}}, 
A.E.Kovalsky\footnote{Moscow Institute of Physics and Technology,
Dolgoprudny, Moscow region. },
A.K.Likhoded\footnote{E-mail: likhoded@mx.ihep.su}.\\
\vspace*{3mm}
Russian State Research Center "Institute for High Energy Physics", Protvino,
Moscow Region, 142284, Russia.
\end{center}
\vspace*{2mm}
\begin{abstract}
In the framework of three-point QCD sum rules, the form factors for the
semileptonic decays of $B_c^+\rightarrow B_s(B_s^*) l^+\nu_l$ are calculated
with account for the Coulomb-like $\alpha_s/v$-corrections in the heavy
quarkonium. The generalized relations due to the spin symmetry of HQET/NRQCD
for the form factors are derived at the recoil momentum close to zero. The
nonleptonic decays are studied using the assumption on the factorization. The
$B_c$ meson lifetime is estimated by summing up the dominating exclusive modes
in the $c \rightarrow s$ transition combining the current calculations with the
previous analysis of $b \rightarrow c$ decays in the sum rules of QCD and
NRQCD.
\end{abstract}

\vspace*{1cm}

\vspace*{1cm}

\section{Introduction}
For better understanding and precise measuring the weak-action properties of
heavy quarks, governed by the QCD forces, we need as wide as possible
collection of snapshots with hadrons, containing the heavy quarks. Then we can
provide the study of heavy quarks dynamics by testing the various conditions,
determining the forming of bound states as well as the entering of strong
interactions into the weak processes. So, a new lab for such investigations is
a doubly heavy long-lived quarkonium $B_c$ recently observed by the CDF
Collaboration \cite{cdf} for the first time. 

This meson is similar to the charmonium and bottomonium in the spectroscopy,
since it is composed by two nonrelativistic heavy quarks, so that the NRQCD
approach \cite{NRQCD} is well justified to the system. The modern predictions
for the mass spectra of $\bar b c$ levels were obtained in refs. \cite{eichten}
in the framework of potential models and lattice simulations. The arrangement
of excitations is close to what was observed in the charmonium and bottomonium.
However, the feature of $B_c$-mesons is an absence of annihilation into light
quarks, gluons and leptons due to QCD and QED, that implies the higher
excitations decay into the low lying levels and  ground state due to the
emission of photons and pion pairs. The measured value of $B_c$ mass yet has a
large uncertainty
$$
M_{B_c} = 6.40\pm 0.39\pm 0.13\; {\rm GeV,}
$$
in agreement with the theoretical expectations.

The production mechanism for the $B_c$-meson was studied in refs. \cite{prod}.
The most simple picture takes place for the production in the
$e^+e^-$-annihilation, where the universal perturbative fragmentation functions
can be analytically calculated for the S-, P- and D-wave levels in the
framework of factorization for the hard production of quarks and their soft
binding into the hadron, which can be reliably described in the potential
models. In hadron collisions, the fragmentation regime takes place at the
transverse momenta $p_T\gg m_{B_c}$, and at $p_T\sim m_{B_c}$ the subleading
terms in $1/p_T$ or higher twists have to be taken into account. This can be
calculated in the framework of factorization approach by a careful evaluation
of complete set of diagrams in the given $\alpha_s$-order, $O(\alpha_s^4)$. The
non-fragmentational contributions dominate at $p_T\sim m_{B_c}$ \cite{prod}. 

The measured $B_c$ lifetime
$$
\tau[B_c] = 0.46^{+0.18}_{-0.16}\pm 0.03\; {\rm ps,}
$$
agrees with the estimates obtained in the framework of both the OPE combined
with the NRQCD evaluation of hadronic matrix elements
\cite{Bigi,Buchalla,BcOnish}
and potential quark models, where one has to sum  up the dominating exclusive
modes
to calculate the total $B_c$ width \cite{LusMas,Bcstatus}
$$
\tau_{\rm OPE,PM}[B_c] = 0.55\pm 0.15\; {\rm ps.}
$$
The accurate measurement of $B_c$ lifetime could allow one to distinguish
various parameter dependencies such as the optimal heavy quark masses, which
basically determine the theoretical uncertainties in OPE. 

At present, the calculations of $B_c$ decays in the framework of QCD sum rules
were performed in \cite{Colangelo,Bagan,KT,Onish}. The authors of
\cite{Colangelo,Bagan}
got the results, where the form factors are about 3 times less than the values
expected in the potential quark models, and the semileptonic and hadronic
widths of $B_c$ are one order of magnitude less than those in OPE. The reason
for such the disagreement was pointed out in \cite{KT} and studied in
\cite{Onish}: in the QCD sum rules for the heavy quarkonia the Coulomb-like
corrections are significant, since they correspond to summing up the ladder
diagrams, where $\alpha_s/v$ is not a small parameter, as the heavy quarks move
nonrelativistically, $v\ll 1$. The Coulomb rescaling of quark-quarkonium vertex
enhances the estimates of form factors in the QCD sum rules for the $B_c^+\to
\psi(\eta_c) l^+ \nu$ decays, where the initial and recoil mesons are both the
heavy quarkonia. In the framework of NRQCD at the recoil momentum close to zero
one derives the spin symmetry relations for the form factors of semileptonic
$B_c$ decays \cite{Jenkins,Onish}. In the strict limit of $v_1=v_2$, where
$v_{1,2}$ denote the four-velocities of initial and recoil mesons,
respectively, the authors of \cite{Jenkins} found a single relation between the
form factors \footnote{In refs.\cite{CF,CH} the relations were studied in the
framework of potential models.}. In \cite{Onish} the soft limit $v_1\cdot
v_2\to 1$ at $v_1\neq v_2$ was considered, and the generalized spin symmetry
relations were obtained for the $B_c\to\psi(\eta_c)$ transitions: four
equations, including that of \cite{Jenkins}. Moreover, the gluon condensate
term was calculated in both QCD and NRQCD, so that it enforced a convergency of
the method.

In the present paper we calculate the $B_c$ decays due to the $c\to s$ weak
transition in the framework of QCD sum rules, taking into account the
Coulomb-like $\alpha_s/v$-corrections for the heavy quarkonium in the initial
state. In the semileptonic decays the hadronic final state is saturated by the
pseudoscalar $B_s$ and vector $B_s^*$ mesons, so that we need the values of
their leptonic constants entering the sum rules and determining the
normalization of form factors. For this purpose, we reanalyze the two-point sum
rules for the $B$ mesons to take into account the product of quark and gluon
condensates in addition to the previous consideration of terms with the quark
and mixed condensates. We demonstrate the significant role of the product term
for the convergency of method and reevaluate the constants $f_B$ as well as
$f_{B_s}$. Taking into account the dependence on the threshold energy $E_c$ of
hadronic continuum in the $\bar b s$ system in both the value of $f_{B_s}$
extracted from the two-point sum rules and the form factors in the three-point
sum rules, we observe the stability of form factors versus $E_c$, which
indicates the convergency of sum rules.

The spin symmetries of leading terms in the lagrangians of HQET \cite{HQET} for
the singly heavy hadrons (here $B_s^{(*)}$) and NRQCD \cite{NRQCD} for the
doubly heavy mesons (here $B_c$) result in the relations between the form
factors of semileptonic $B_c\to B_s^{(*)}$ decays. We derive two generalized
relations in the soft limit $v_1\cdot v_2\to 1$: one equation in addition to
what was found previously in ref.\cite{Jenkins}. The relations are in a good
agreement with the sum rules calculations up to the accuracy better than 10\%,
that shows a low contribution of next-to-leading $1/m_Q$-terms.

We perform the numerical estimates of semileptonic $B_c$ widths and use the
factorization approach \cite{blokshif} to evaluate the hadronic modes.
Summing up the dominating exclusive modes, we calculate the lifetime of $B_c$,
which agree with the experimental data and the predictions of OPE and quark
models. We discuss the preferable prescription for the normalization point of
nonleptonic weak lagrangian for the charmed quark and present our optimal
estimate of total $B_c$ width. We stress that in the QCD sum rules to the given
order in $\alpha_s$, the uncertainty in the values of heavy quark masses is
much less than in OPE. This fact leads to a more definite prediction on the
$B_c$ lifetime.

The paper is organized as follows. Section 2 is devoted to the general
formulation of three-point sum rules for the $B_c$ decays with account of
Coulomb-like corrections. The analysis of two-point sum rules for the leptonic
constant of singly heavy meson with the introduction of term allowing for the
product of quark and gluon condensates is presented in Section 3, where we
also show the convergency of three-point sum rules with respect to a dependence
on the threshold energy of continuum in the heavy-light system. We estimate the
form factors of semileptonic $B_c\to B_s^{(*)}$ decays. The relations between
the form factors of semileptonic decays as follows from the spin symmetry of
HQET and NRQCD are derived in Section 4 in the soft limit of zero recoil.
Section 5 contains the description how the nonleptonic decays modes are
calculated and the $B_c$ lifetime is evaluated. We discuss the optimal
estimation of lifetimes for the heavy hadrons in Section 6. In Conclusion we
summarize the results.

\section{Three-point sum rules for the $B_c$ meson.}
In this paper the approach of three-point QCD sum rules \cite{Shif} is used to
study the form factors of semileptonic and nonleptonic decay rates for the c
$\rightarrow$ s transition in decays of $B_c$ meson. From the  two-point sum
rules we extract the values for the leptonic constants of mesons in the initial
and final states. In our consideration we use the following notations:

\begin{equation}
 \langle 0|\bar q_1 i \gamma_5 q_2|P(p) \rangle=\frac{f_P M^2_P}{m_1+m_2},\  
\label{Lepconps}
\end{equation}
\
 and
\begin{equation}
\langle 0|\bar q_1 i \gamma_{\mu} q_2|V(p,\epsilon)\rangle=i \epsilon_{\mu} M_V
f_V   ,
\label{Lepconv}
\end{equation}
where P and V represent the scalar and vector mesons, and $m_1$, $m_2$ are
the quark masses.

The hadronic matrix elements for the semileptonic $c \rightarrow s~$transition
in the $B_c$ decays can be written down as follows: 
\begin{eqnarray}
\langle B_s(p_2)|V_{\mu}|B_c(p_1)\rangle &=& f_{+}(p_1 + p_2)_{\mu} +
f_{-}q_{\mu},\\
\frac{1}{i}\langle B_s^* (p_2)|V_{\mu}|B_c(p_1)\rangle &=& 
i F_V\epsilon_{\mu\nu\alpha\beta}\epsilon^{*\nu}(p_1 +
p_2)^{\alpha}q^{\beta},\\
\frac{1}{i}\langle B_s^* (p_2)|A_{\mu}|B_c(p_1)\rangle &=&
F_0^A\epsilon_{\mu}^{*} + 
F_{+}^{A}(\epsilon^{*}\cdot p_1)(p_1 + p_2)_{\mu} + 
F_{-}^{A}(\epsilon^{*}\cdot p_1)q_{\mu}, 
\end{eqnarray}
where $q_{\mu} = (p_1 - p_2)_{\mu}$ and $\epsilon^{\mu} = \epsilon^{\mu}(p_2)$
is the polarization vector of $B_s^*$ meson. $V_{\mu}$ and $A_{\mu}$ are the
flavour changing vector and axial electroweak currents. The form factors
$f_{\pm}, F_V, F_0^A$ and $F_{\pm}^A$ are functions of $q^2$ only. It should be
noted that since the leptonic current $l_{\mu} = \bar l\gamma_{\mu}(1 +
\gamma_5)\nu_l$ is transversal in the limit of massless leptons, the
probabilities of semileptonic decays are independent of $f_-$ and $F_-^A$ (the
$\tau^{+} \nu_{\tau}$ mode is forbidden by the energy conservation). Following
the standard procedure for the evaluation of form factors in the framework of
QCD sum rules \cite{SR3pt}, we consider the three-point functions
\begin{eqnarray}
\Pi_{\mu}(p_1, p_2, q^2) &=& i^2 \int dxdye^{i(p_2\cdot x - p_1\cdot
y)} \cdot \nonumber\\
&&\langle 0|T\{\bar q_1(x)\gamma_5 q_2(x), V_{\mu}(0), \bar b(y)\gamma_5
c(y)\}|0
\rangle,\\
\Pi_{\mu\nu}^{V, A}(p_1, p_2, q^2) &=& i^2 \int dxdye^{i(p_2\cdot x - p_1\cdot
y)} \cdot \nonumber\\
&&  \langle 0|T\{\bar q_1(x)\gamma_{\mu} q_2(x), J_{\nu}^{V, A}(0), 
\bar b(y)\gamma_5 c(y)\}|0\rangle,
\end{eqnarray}
where $\bar q_1(x)\gamma_5 q_2(x)$ and $\bar q_1(x)\gamma_{\nu}q_2(x)$ denote
interpolating
currents for $B_{s}$ and $B_{s}^*$, correspondingly. 
$J_{\mu}^{V, A}$ are the currents $V_{\mu}$ and $A_{\mu}$ of relevance to the
various cases.

The Lorentz structures in the correlators can be written down as
\begin{eqnarray}
\Pi_{\mu} &=& \Pi_{+}(p_1 + p_2)_{\mu} + \Pi_{-}q_{\mu},\label{R1}\\
\Pi_{\mu\nu}^V &=& i\Pi_V\epsilon_{\mu\nu\alpha\beta}p_2^{\alpha}p_1^{\beta},\\
\Pi_{\mu\nu}^A &=& \Pi_{0}^{A}g_{\mu\nu} + \Pi_{1}^{A}p_{2, \mu}p_{1, \nu} +
\Pi_{2}^{A}p_{1, \mu}p_{1, \nu} + \Pi_{3}^{A}p_{2, \mu}p_{2, \nu} + 
\Pi_{4}^{A}p_{1, \mu}p_{2, \nu}.
\end{eqnarray}
\noindent
The form factors $f_{\pm}$, $F_V$, $F_{0}^{A}$ and $F_{\pm}^{A}$ are determined
from the amplitudes $\Pi_{\pm}$, $\Pi_V$, $\Pi_{0}^{A}$ and $\Pi_{\pm}^{A} =
\frac{1}{2}(\Pi_{1}^{A}\pm \Pi_{2}^{A})$, respectively. In (8)-(10) the scalar
amplitudes $\Pi_i$ are the functions of kinematical invariants, i.e. $\Pi_i =
\Pi_i(p_1^2, p_2^2, q^2)$.

The leading QCD term is a triangle quark-loop diagram in Fig. \ref{Diag-fig},
for which we can write down the double dispersion representation at $q^2\leq
0$
\begin{equation}
\Pi_i^{pert}(p_1^2, p_2^2, q^2) = -\frac{1}{(2\pi)^2}\int
\frac{\rho_i^{pert}(s_1, s_2, Q^2)}{(s_1 - p_1^2)(s_2 - p_2^2)}ds_1ds_2 + 
\mbox{subtractions},
\label{pertdisp}
\end{equation}
where $Q^2 = -q^2 \geq 0$. The integration region in (\ref{pertdisp}) is
determined by the condition
\begin{equation}
-1 < \frac{2s_1s_2 + (s_1 + s_2 - q^2)(m_b^2 - m_c^2 - s_1)}
{\lambda^{1/2}(s_1, s_2, q^2)\lambda^{1/2}(m_c^2, s_1, m_b^2)} < 1,
\end{equation}
where $\lambda(x_1, x_2, x_3) = (x_1 + x_2 - x_3)^2 - 4x_1x_2$. 
The expressions
for spectral densities $\rho_i^{pert}(s_1, s_2, Q^2)$ are given in Appendix A.

\setlength{\unitlength}{1mm}
\begin{figure}[th]
\begin{center}
\begin{picture}(100, 60)
\put(0, 0){\epsfxsize=11cm \epsfbox{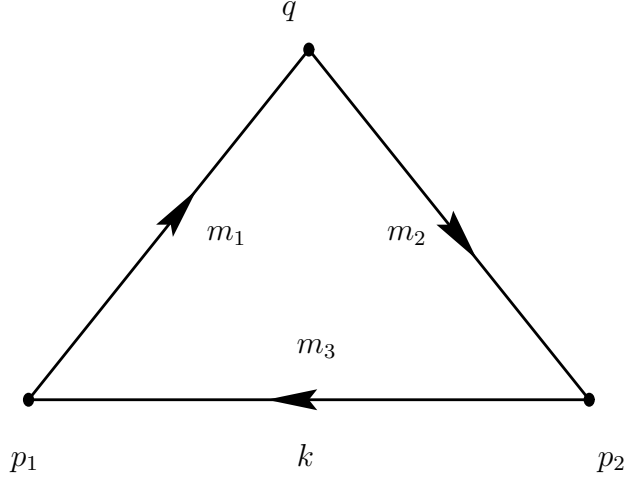}}
\put(85, 10){$p_{2}$}
\put(43,70){$q$}
\put(7, 10){$p_{1}$}
\put(45, 10){$k$}
\put(45, 25){$m_3$}
\put(33, 40){$m_1$}
\put(57, 40){$m_2$}
\end{picture}
\end{center}
\caption{The triangle diagram, giving the leading perturbative term in the OPE
expansion of three-point function.} 
\label{Diag-fig}
\end{figure}

Now let us proceed with the physical part of three-point sum rules. The
connection to hadrons in the framework of QCD sum rules is obtained by matching
the resulting QCD expressions of current correlators with the spectral
representation, derived from a double dispersion relation at $q^2\leq 0$.
\begin{equation}
\Pi_i(p_1^2, p_2^2, q^2) = -\frac{1}{(2\pi)^2}\int
\frac{\rho_i^{phys}(s_1, s_2, Q^2)}{(s_1 - p_1^2)(s_2 - p_2^2)}ds_1ds_2 + 
\mbox{subtractions}.
\label{physdisp}
\end{equation}
Assuming that the dispersion relation (\ref{physdisp}) is well convergent, the
physical spectral functions are generally saturated by the ground 
hadronic states and a continuum starting at some effective thresholds
$s_1^{th}$ and $s_2^{th}$
\begin{eqnarray}
\rho_i^{phys}(s_1, s_2, Q^2) &=& \rho_i^{res}(s_1, s_2, Q^2) + \\
&& \theta (s_1-s_1^{th})\cdot\theta (s_2-s_2^{th})\cdot
\rho_i^{cont}(s_1, s_2, Q^2),
\nonumber
\end{eqnarray}
where the resonance term is expressed through the product of leptonic constant
and form factor for the transition under consideration, so that
\begin{eqnarray}
\rho_i^{res}(s_1, s_2, Q^2) &=& {\langle 0| \bar b\gamma_{5}(\gamma_{\mu})
s|B_s(B^*_s) \rangle\; \langle B_s(B^*_s)|F_i(Q^2)|B_c\rangle\; \langle
B_c|\bar b\gamma_5 c|0)\rangle}\cdot \nonumber\\
&& {(2\pi)^2 \delta(s_1-M_1^2) \delta(s_2-M_2^2)}
+ \mbox{higher~state~contributions},
\end{eqnarray}
where $M_{1,2}$ denote the masses of hadrons in the initial and final states.
The continuum of higher states is modelled by the perturbative absorptive part
of $\Pi_i$, i.e. by $\rho_i$.  Then, the expressions for the form factors $F_i$
can be derived by equating the representations for the three-point functions
$\Pi_i$ in (\ref{pertdisp}) and (\ref{physdisp}), which means the formulation
of sum rules.
  
For the heavy quarkonium $\bar b c$, where the relative velocity of quark
movement is small, an essential role is taken by the Coulomb-like
$\alpha_s/v$-corrections. They are caused by the ladder diagram, shown in
Fig. \ref{Coul-fig}. It is well known that an account for this corrections in
two-point sum rules numerically leads to a double-triple multiplication of Born
value of spectral density \cite{Shif,fbc}. In our case it leads to the finite
renormalization for $\rho_i$ \cite{Onish}, so that
\begin{equation}
 \rho^{c}_i={\bf C} \rho_i,
\label{ren}
\end{equation}

\setlength{\unitlength}{1mm}
\begin{figure}[th]
\begin{center}
\begin{picture}(110,100)
\put(0, 0){\epsfxsize=11cm \epsfbox{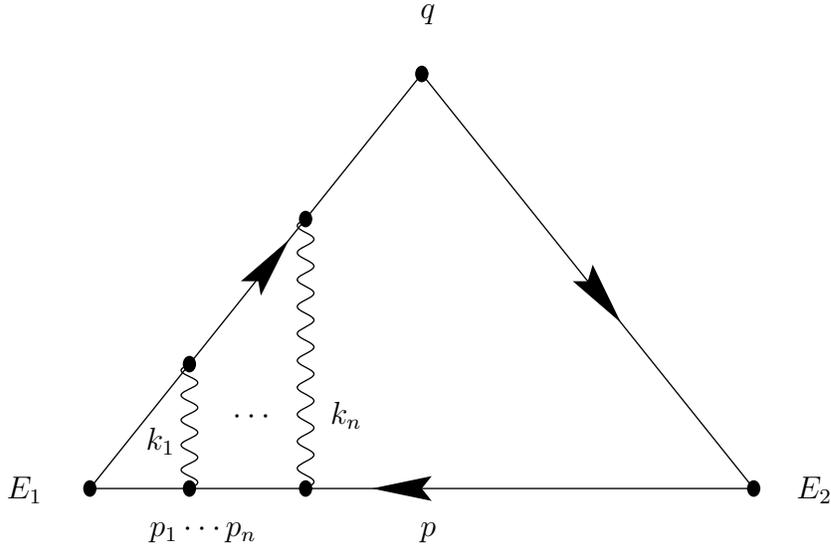}}
\put(0,40){$E_1$}
\put(105,40){$E_2$}
\put(55,35){$p$}
\put(55,104){$q$}
\put(19,35){$p_1\cdots p_n$}
\put(30,50){$\cdots$}
\put(18.5,46.5){$k_1$}
\put(43,50){$k_n$}
\end{picture}
\end{center}
\normalsize

\vspace*{-3cm}
\caption{The ladder diagram of the Coulomb-like interaction.} 
\label{Coul-fig}
\end{figure}

\begin{equation}
 {\bf C}=\frac{|\Psi^C_{\bar b c}(0)|}{|\Psi^{free}_{\bar b
 c}(0)|}=\sqrt{\frac{4\pi
 \alpha_s}{3v}(1-\exp\{-\frac{4\pi \alpha_s}{3v}\})^{-1}},
\end{equation}
where $v$ is the relative velocity of quarks in the $\bar b c$-system,
\begin{equation}
v=\sqrt{1-\frac{4 m_b m_c}{p_1^2-(m_b-m_c)^2}}.
\end{equation}
To the moment, the procedure of calculations is completely described.

\section{Numerical results on the form factors and the semileptonic decay
widths}\
We evaluate the form factors in the scheme of spectral density moments. This
scheme is not strongly sensitive to the value of the $\bar b c$-system
threshold energy. In our calculations $E_c^{\bar b c}=1.2~\mbox{GeV}$. The
two-point sum rules for the $B_c$ meson with account for the Coulomb-like
corrections give $\alpha_s^{c}(\bar b c)$=0.45, which corresponds to
$f_{B_c}$=400 MeV \cite{fbc}. The quark masses are fixed by the calculations of
leptonic constants $f_{\Psi}$ and $f_{\Upsilon}$ in the same order over
$\alpha_s$. The requirement of stability in the sum rules  including the
contributions of higher excitations, results in quite an accurate determination
of masses $m_c=1.40\pm0.03~$GeV and $m_b=4.60\pm0.02~$GeV, which are in a good
agreement with the recent estimates in \cite{masses}, where the quark masses
free off a renormalon ambiguity were introduced. The values of leptonic
constants $f_{\Psi}$, $f_{\Upsilon}$ linearly depend on the Coulomb-exchange
$\alpha_s$. We find $\alpha_s^c(\bar c c)\simeq0.60$, $\alpha_s^c(\bar b
b)\simeq0.37$, which obey the remormalization group evolution with the
appropriate scale prescription, depending on the quarkonium contents. In this
way, we can extract the above values of $\alpha_s^c(\bar b c)$ and $f_{B_c}$.
Note, that the heavy ($Q_1 \bar Q_2$)-quarkonia constants obey the scaling
relation \cite{fbc,Bcstatus}
\begin{equation}
 \frac{f_n^2}{M_n}\left(\frac{M_n(m_1+m_2)}{4 m_1 m_2}\right)^2=\frac{c}{n},
 \end{equation}
where $n$ denotes the radial excitation number of nS-level, and $c$ is
independent of heavy quark flavors.
 
The leptonic constant for the $B_s$ meson is extracted from the two-point sum
rules. The Borel improved sum rules for the $B$ meson leptonic constant
\cite{Neubert} have the following form:
\begin{equation}
 f^2_B M_Be^{-\bar \Lambda(\mu) \tau}=K^2 \frac{3}{\pi^2} C(\mu)
 \int\limits^{\omega_0(\mu)}_{0}
 d\omega~\omega^2 e^{-\omega \tau}+\langle \bar q q \rangle(1-
 \frac{m_0^2~\tau^2}{16}+ \frac{\pi^2 \tau^4}{288}\langle\frac{\alpha_s}{\pi}
 G^2 \rangle),
\label{Tcond} 
\end{equation}
where we use $\langle \bar q q \rangle=-(0.23\cdot \mbox{GeV})^3$,
$m_0^2=0.8~\mbox{GeV}^2$, $\langle\frac{\alpha_s}{\pi} G^2
\rangle=1.77\cdot10^{-2}~\mbox{GeV}^4$ as the central values, and $M_B$=5.28
GeV. The K-factor is due to $\alpha_s$-corrections. We expect it is large, but
we suppose the appearance of the same factor in evaluating the
$\alpha_s$-corrections to the heavy-light vertex in the triangle diagram. For
this factor we have the following expression \cite{Neubert}:
\begin{equation}
K^2=\biggl\{ \int \limits_{0}^{E_c \tau}z^2 e^{-z} dz \biggr\}^{-1}\cdot \int
\limits_{0}^{E_c \tau}z^2 e^{-z} \biggl\{1+\frac{2 \alpha_s(\tilde
\Lambda)}{\pi}\left(\frac{13}{6}+\frac{2 \pi^2}{9}-\mbox{ln}~ z\right)
\biggr\}dz,  
\label{Kfactor}
\end{equation}
where we suppose that the scale $\tilde \Lambda$ is equal to 1.25 GeV. The
dependence of K-factor on the Borel parameter $\tau$ and the threshold energy
$E_c$ is shown in Fig. \ref{K3d-fig}. The K-factor is not
sensitive to $E_c$ changing in the range $1.0\div1.5~$GeV.

\setlength{\unitlength}{1mm}
\begin{figure}[th]
\begin{center}
\begin{picture}(100, 80)
\put(0,0){\epsfxsize=11cm \epsfbox{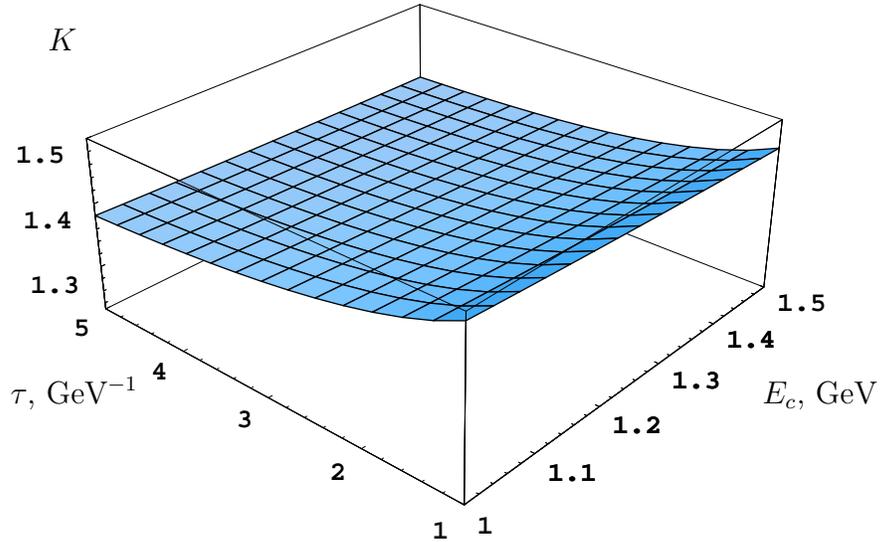}}
\put(0, 23){$\tau$,$~\mbox{GeV}^{-1}$}
\put(5, 70){$K$}
\put(100, 23){$E_c$, GeV}
\end{picture}
\end{center}
\caption{The K-factor dependence on the Borel parameter $\tau$ and the
threshold
energy $E_c$.} 
\label{K3d-fig}
\end{figure}

We see that NLO corrections to the leptonic constant are about $40\%$. Using
the Pad\'e approximation, we find that higher orders corrections can be about
$30\%$. So, we hold the K factor in conservative limits $1.4\div1.7$. It is
quite reasonable to suppose its cancellation in evaluating the semileptonic
form factors due to the renormalization of heavy-light vertex in the triangle
diagram.

The contribution of quark condensate term is not sensitive to the variation of
$\langle \bar q q \rangle$ in the limits from -$(0.23\cdot\mbox{GeV})^3$ to
-$(0.27\cdot\mbox{GeV})^3$ (this variation corresponds to the renormalization
group evolution and the insertion of $\alpha_s$-corrections to this term, so
that $K$ can be putted as the overall factor). 

In the limit of semi-local duality \cite{fesr,SU(3)} $\tau \rightarrow 0$ we
get the relation: $\bar \Lambda(\mu) = \frac{3}{4}~\omega_0(\mu)$ (the
contribution of the quark condensate term to this equation is about $\sim 15\%$
). We introduce the renormalization invariant quantities
$$
\omega_{0,dual}^{ren}=C^{-1/3}(\mu)~\omega_0(\mu),\hspace*{11mm}
\bar\Lambda^{ren}_{dual} = \frac{3}{4}~\omega_{0,dual}^{ren}.
$$ 
For $\bar\Lambda^{ren}_{dual}$ we have $\bar \Lambda^{ren}_{dual} = M_B-m_b =
0.63~\mbox{GeV}$, and we obtain that in the semi-local duality the threshold
energy $~\omega_{0,dual}^{ren}=0.84~\mbox{GeV}$. Neglecting the quark
condensate term in the leptonic constant we have 
\begin{equation} 
f_B^2M_B=K^2 \frac{3}{\pi^2}(\omega_{0,dual}^{ren})^3.
\end{equation}
Since in the three-point sum rules we use the scheme of moments and search for
a stable region, in the general Borel scheme for $f_B$ we have to consider the
stability at $\tau \neq0$ with the extended region of resonance contribution.
We expect, that the sum rules in (\ref{Tcond}) with the redefined
$\omega^{ren}$ and $\bar \Lambda^{ren}$, as mentioned, have a stability point
at $\tau \sim \frac{1}{\bar \Lambda}$
\begin{equation}
f_B^2M_B e^{-\bar \Lambda \tau}=K^2
\frac{3}{\pi^2}\int\limits^{E_c}_{0}
 d\omega~\omega^2 e^{-\omega \tau}+\langle \bar q q
\rangle(1-
 \frac{m_0^2~\tau^2}{16}+ \frac{\pi^2 \tau^4}{288}\langle
 \frac{\alpha_s}{\pi}G^2
 \rangle),
\label{Ecfb}
\end{equation}
where $E_c$ is already not equal to $\omega_{0,dual}^{ren}$. Demanding a low
deviation of $\bar \Lambda$ from $\bar \Lambda^{ren}=0.63$ GeV, we find that
sum rules in Eq.(\ref{Ecfb}) can lead to the results, which are in a good
agreement with the semi-local duality if $E_c=1.1\div1.3~$GeV (see
Fig. \ref{Tcond-fig}). Then the optimal value of Borel parameter $\tau=\tau_m
\simeq 6.5~ \mbox{GeV}^{-1}$. We write down 
\begin{equation}
f_B^2M_B e^{-\bar\Lambda \tau_m}=K^2 \frac{3}{\pi^2}RE_c^3+\langle \bar q q
\rangle(1-
 \frac{m_0^2~\tau^2_m}{16}+ \frac{\pi^2 \tau^4_m}{288}\langle
 \frac{\alpha_s}{\pi}
 G^2 \rangle),
\label{fB} 
\end{equation}
where $R$ denotes the average value of $e^{-\omega \tau_m}$. So, we find the
$E_c^{3/2}$-dependence of $f_B \sqrt{M_B}$, whereas the contribution of
condensate is numerically suppressed, as expected from the semi-local duality.
The results of general Borel scheme calculations of $f_B \sqrt{M_B}$ ignoring
the overall $K$-factor are presented in Fig. \ref{Tcond-fig}. We observe two
stability regions. The stability region at $\tau=2 \div 4$ corresponds to that
of considered in \cite{Neubert}. The results for the leptonic constant $f_B$
obtained from this region \cite{Neubert} is about 1.5 greater than the value
obtained from the stability region at $\tau=6\div7$. The second region appears
only when we introduce the term with the product of quark and gluon
condensates. The similar situation has been observed in the NRQCD sum rules for
doubly heavy baryons \cite{Onishbar}. The product of quark and gluon
condensates was not taken into account in \cite{Neubert}, and therefore, the
intermediate stability point was observed only. Fixing the optimal values of
$f_B^2 M_B$ in Eq.(\ref{fB}) from Fig. \ref{Tcond-fig}, we can invert the sum
rules to study the dependence of $\bar \Lambda$ on $\tau$, as shown in
Fig. \ref{Lambda-fig}, where the optimal values of $\bar\Lambda$ agree with the
semi-local duality and the estimate $\bar\Lambda^{ren}=M_B-m_b$. Note that the
intermediate stability point $\tau \sim 4$ exhibit a low variation of $\bar
\Lambda$ close to 0.4 GeV, which was obtained in \cite{Onishbar,Neubert}, and
usually given by the potential models (see, for instance, \cite{Spbar}). 

\setlength{\unitlength}{1mm}
\begin{figure}[ph]
\begin{center}
\begin{picture}(100, 73)
\put(0, 0){\epsfxsize=11cm \epsfbox{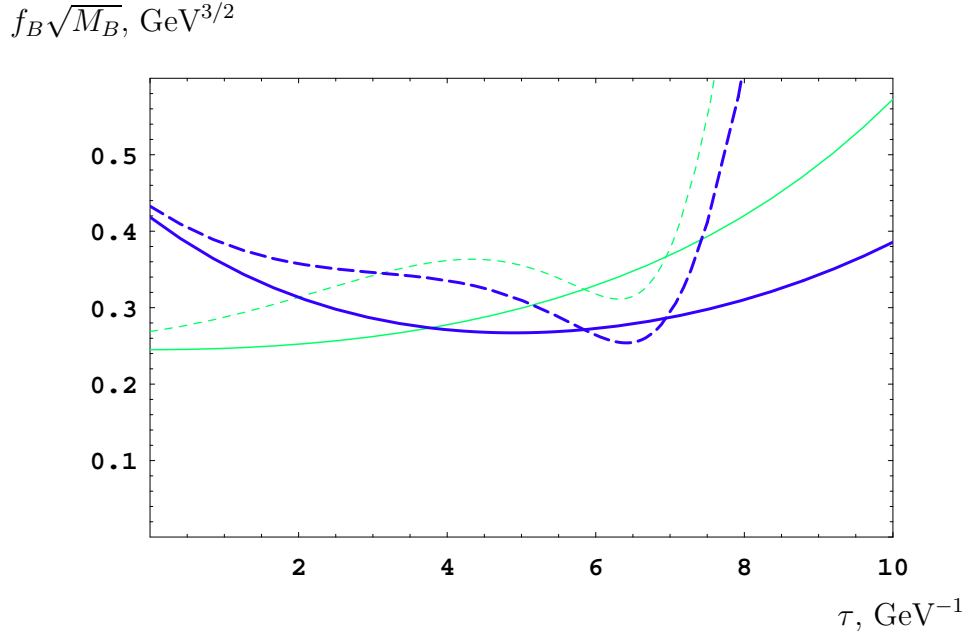}}
\put(100, -5){$\tau$,$~\mbox{GeV}^{-1}$}
\put(-10, 74){$f_B \sqrt{M_B}$,$~\mbox{GeV}^{3/2}$}
\end{picture}
\end{center}

\vspace*{4mm}
\caption{ $f_{B}$ in the semi-local duality sum rules (pale curves) and in the
general Borel scheme. Solid lines correspond to the sum rules without
condensate terms, dashed lines correspond to the accounting for the condensates
contributions.} 
\label{Tcond-fig}
\end{figure}

\begin{figure}[ph]
\begin{center}
\begin{picture}(100, 60)
\put(0, 0){\epsfxsize=11cm \epsfbox{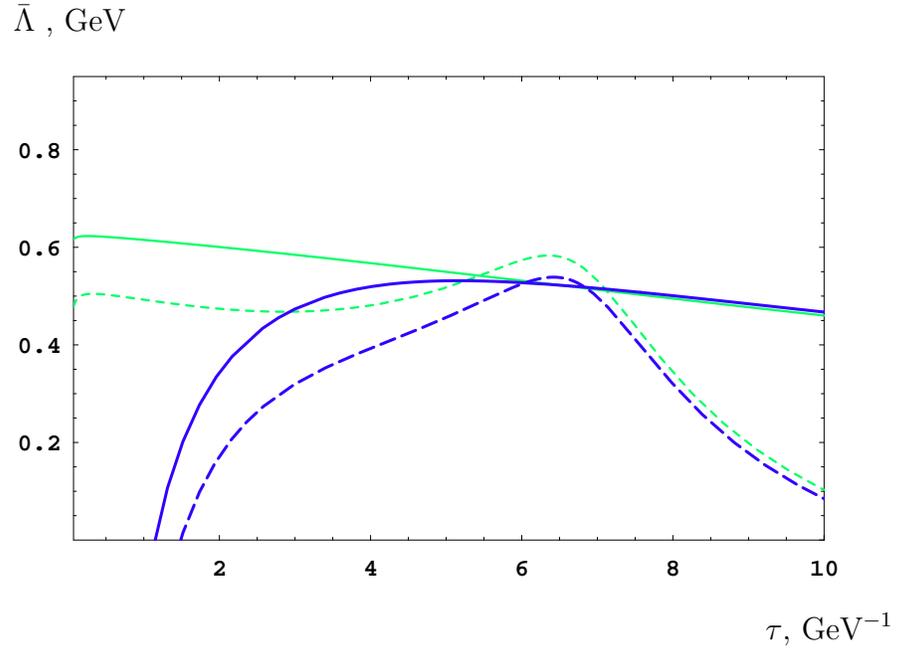}}
\put(100, -7){$\tau$,$~\mbox{GeV}^{-1}$}
\put(0, 74){$\bar \Lambda~$, GeV}
\end{picture}
\end{center}

\vspace*{4mm}
\caption{ The dependence of $\bar \Lambda$ on $\tau$ with the fixed values of
$f_B \sqrt{M_B}$, which correspond to the stability regions at $\tau=6\div7$
GeV$^{-1}$ (The notations are the same as in Fig. \ref{Tcond-fig}).} 
\label{Lambda-fig}
\end{figure}

The calculated physical quantity should be independent of
parameters in the sum rule scheme. However, we truncate both the operator
product expansion and the perturbative series for the Wilson coefficients by
fixed orders in the opreator dimension and $\alpha_s$, respectively. This fact
leads to that the results depend on the scheme parameters, say, the Borel
variable. Moreover, the physical part of sum rules is modelled by the
contributions of resonances and a continuum term starting at a threshold, that
introduces the dependence on the threshold value and suggests that the
stability of results can be improved by the terms of excited resonaces in
addition to the ground state.

\begin{figure}[th]
\begin{center}
\begin{picture}(100, 73)
\put(0, 0){\epsfxsize=11cm \epsfbox{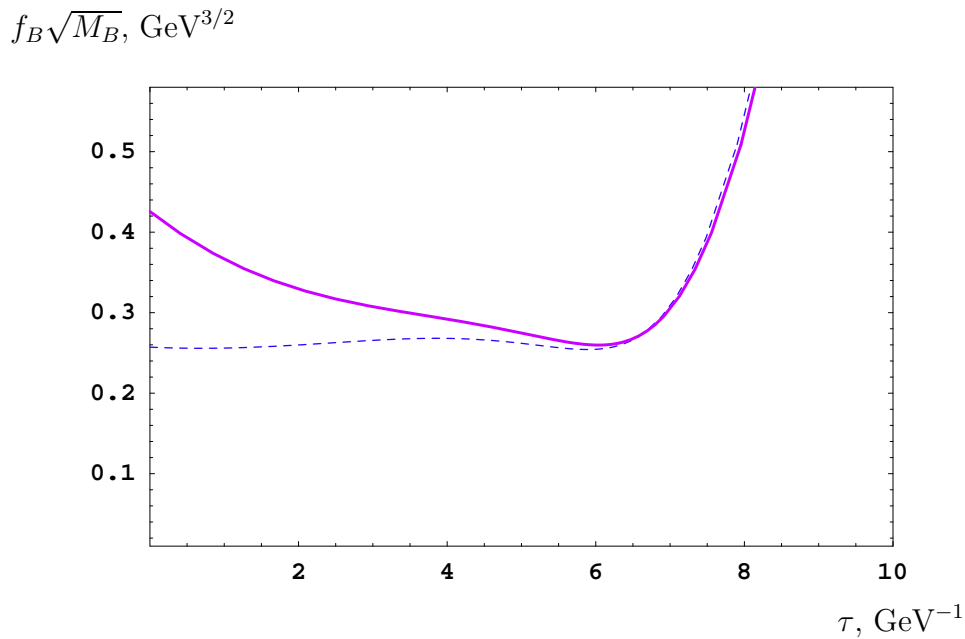}}
\put(100, -5){$\tau$,$~\mbox{GeV}^{-1}$}
\put(-10, 74){$f_B \sqrt{M_B}$,$~\mbox{GeV}^{3/2}$}
\end{picture}
\end{center}

\vspace*{4mm}
\caption{ $f_{B}$ in the semi-local duality sum rules (dashed curve) and in the
general Borel scheme (solid line) with the corrected value of $\bar \Lambda$,
which improves the stability of result obtained in the semi-local duality.} 
\label{stabil}
\end{figure}

Let us, at first, consider the results on the leptonic constant in the limit of
semi-local duality, which requires the stability at $\tau\to 0$ and corresponds
to the duality region containing the ground state, only. The sum rules show
that the condensate contributions are given by the polinomials over $\tau$, and
the leading correction at $\tau\to 0$ is the term with the quark condensate.
Then, we expect that the region of stability will extend at $\tau > 0$ if we
will add the higher condensates with the appropriate ratios of their values.
The central values of condensates as mentioned above correspond to the results
shown in Fig. \ref{Tcond-fig}. The stability of semi-local sum rules can be
improved by a variation of $\bar\Lambda$ and $\omega_0$, which is not important
for the current discussion. In order to clarify this statement we present the
results of semi-local sum rules for the leptonic constant of $B$ meson in Fig.
\ref{stabil}, where we put $\bar \Lambda = 0.54$ GeV. Note, that the value of
leptonic constant is the same as we have found in the pertubative limit of
semi-local duality. Then, we state that the value of leptonic constant obtained
at $\tau\to 0$ agrees with the value corresponding to the stability at $\tau
\approx 7$ GeV$^{-1}$, i.e. in the second point of local extremum in Fig.
\ref{Tcond-fig}. In order to confirm, we present also the result for the other
ratio of condensate values (the lower value of mixed condensate and the upper
value of gluon condensate in the regions mentioned below) in Fig. \ref{Tcond'}.
We see that the semi-local duality is completely broken at unappropriate choice
of condensate values.

\setlength{\unitlength}{1mm}
\begin{figure}[ph]
\begin{center}
\begin{picture}(100, 73)
\put(0, 0){\epsfxsize=11cm \epsfbox{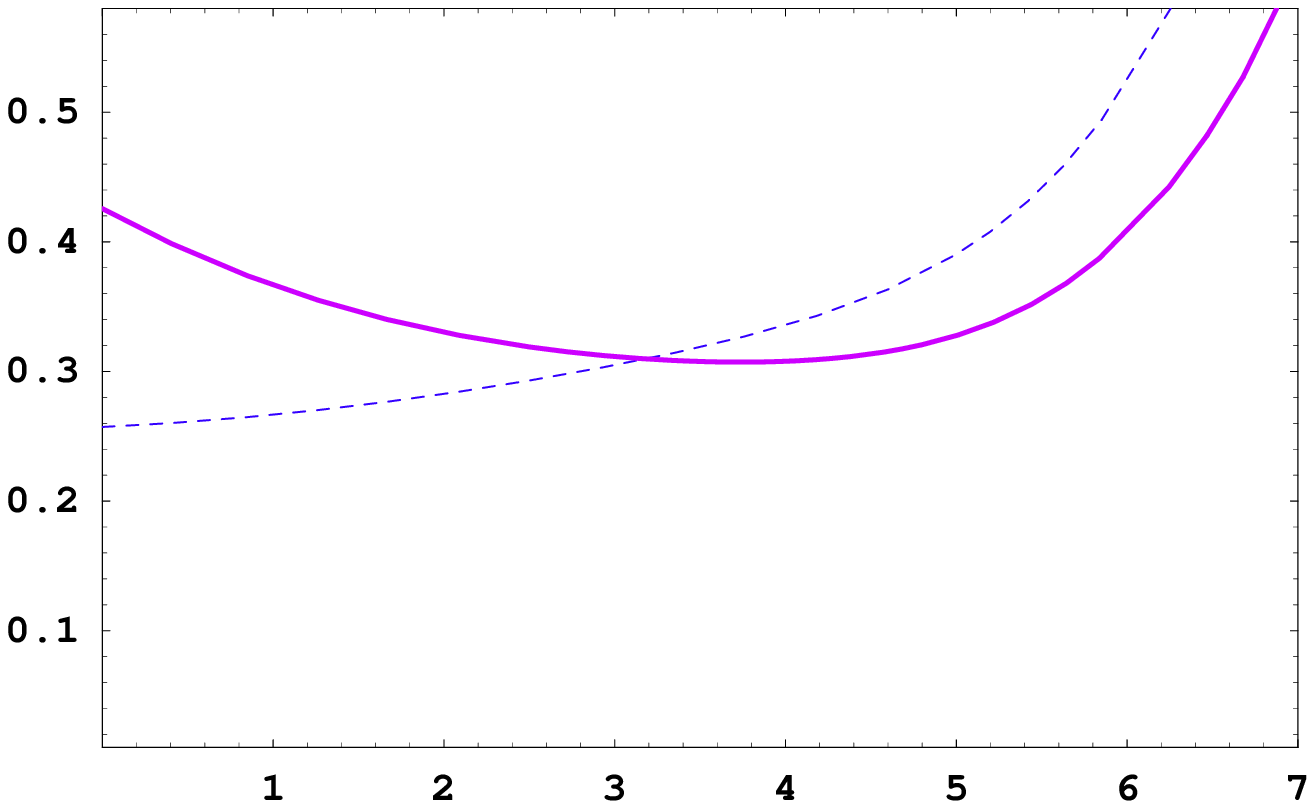}}
\put(100, -5){$\tau$,$~\mbox{GeV}^{-1}$}
\put(-10, 74){$f_B \sqrt{M_B}$,$~\mbox{GeV}^{3/2}$}
\end{picture}
\end{center}

\vspace*{4mm}
\caption{ $f_{B}$ in the semi-local duality sum rules (dashed curve) and in the
general Borel scheme (solid line) with the condensate values, which
destroy the semi-local duality.} 
\label{Tcond'}
\end{figure}


\begin{figure}[ph]
\begin{center}
\begin{picture}(100, 60)
\put(0, 0){\epsfxsize=11cm \epsfbox{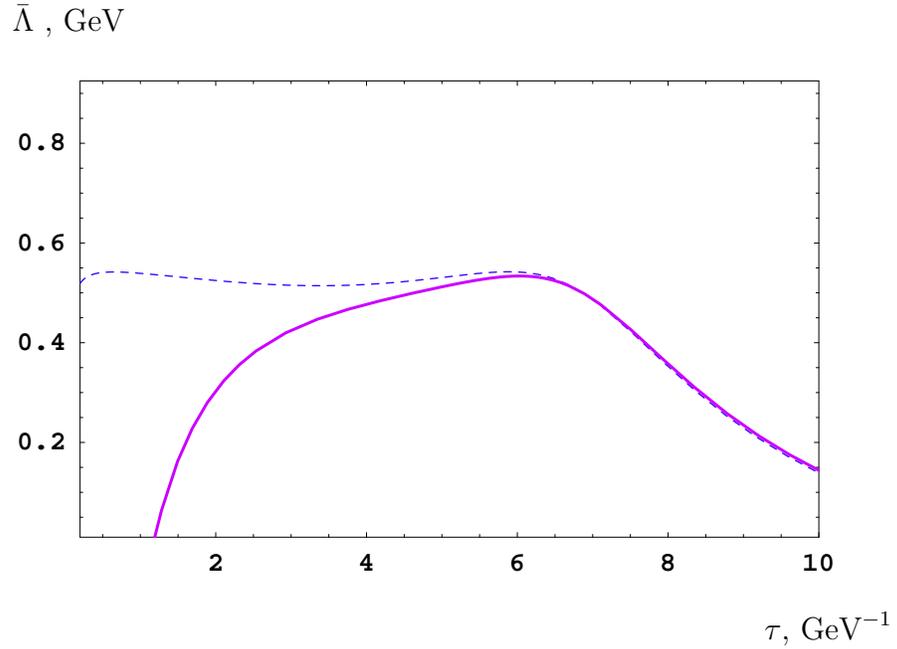}}
\put(100, -7){$\tau$,$~\mbox{GeV}^{-1}$}
\put(0, 74){$\bar \Lambda~$, GeV}
\end{picture}
\end{center}

\vspace*{4mm}
\caption{The dependence of $\bar \Lambda$ on $\tau$ with the fixed values of
$f_B \sqrt{M_B}$, which correspond to the stability regions at $\tau=6\div7$
GeV$^{-1}$ in Fig. \ref{stabil}.} 
\label{Lambda}
\end{figure}

Second, we study the general Borel sum rules in the scheme, where we do not
require the stability at $\tau\to 0$, which means that we extend the duality
region to incorporate possible excitation contributions. Then, we expect that
the estimate of leptonic constant should result in the same value obtained in
the semi-local duality. We see that this can be reached at the same values of
condensates involved as well as in the same region of $\tau$, which is again
given by the second extremum (see Figs. \ref{Tcond-fig} and \ref{stabil}).
Certainly, the extension of duality region leads to a failure of stability at
low $\tau$, which is caused by additional terms given by the excitations beyond
the control of the method. 

At higher values of $\tau$ the stability would be reached
by introduction of higher condensates contributing as the polinomials of higher
powers. Of course, we can achieve the better stability in the general Borel
scheme by adjusting the condensate values, but this will destroy the semi-local
duality, that indicates the divergency of the method, while the convergency
demands an appropriate choice of condensate values as we put. 

The criterion on the covergency of both the semi-local duality and general
Borel sum rules was ignored in \cite{Neubert}, where the greater value of
leptonic constant was obtained (see Fig. \ref{Tcond'}).

The same notes can be done in the discussion on $\bar \Lambda$. For the sake of
comparison, we present the results in Figs. \ref{Lambda-fig} and \ref{Lambda},
corresponding to the inverted sum rules for the leptonic constant given in
Figs. \ref{Tcond-fig} and \ref{stabil}.

The physical meaning of gluon condensate as it stands in the operator
product expansion taken between the observed states is independent of the
scheme of calculations. In the sum rules the gluon condensate contributes in
the region, where the excited states are suppressed, while the ground state
corresponding to the current under consideration dominates. So, the condensate
essentially determines the binding energy of quarks in the hadron containing
the heavy quark. The definition of sum rules scheme includes the parameter
determining the threshold energy of continuum contribution in addition to the
resonance term. So, the difference between the schemes of semi-local duality
and usual Borel representation is due to the variation of duality region. So,
the semi-local duality means the duality for the region contaning the ground
state only, while the usual Borel (or moment) scheme explore the extended
region containing several hadronic states: the ground state and its
excitations. However, the gluon condensate essentially contributes at the
scheme parameters, when the excitations are suppressed. Thus, its value is the
same for both schemes used, i.e. for the semi-local duality and in the Borel
scheme. The value of gluon condensate has been varied in the range
$\langle\frac{\alpha_s}{\pi} G^2 \rangle=(1.5\div 2)\cdot10^{-2}~\mbox{GeV}^4$.
As for the mixed condesate we have used the range $m_0^2=(0.72\div
0.88)~\mbox{GeV}^2$. The variation of these parameters does not change the
qualitative picture for the lepton constant as it has been discussed, while the
numerical uncertainty of its value is less than 7\%.

Numerically, multiplying the result taken from Fig. \ref{Tcond-fig}, by the
K-factor we find the value $f_B=140\div170~$MeV, which is in a good agreement
with the recent lattice results \cite{Lellouch} and the estimates in the QCD SR
by other authors \cite{Braun}. So, we can conclude that the $1/m_b$-corrections
are not valuable for $f_B$. The uncertainty of estimates is basically connected
with the higher orders in $\alpha_s$. For the vector $B^*$ meson constant
$f_{B^*}$ we put $\frac{f_{B^*}}{f_{B}}=1.11~$(see \cite{Neubert2,Braun}). 

For the leptonic constant of $B_s$ meson we explore the following relation
$\frac{f_{B_s}}{f_B}=1.16$, which expresses the flavor SU(3)-symmetry violation
for B mesons \cite{SU(3)}.
 
Remember, in sum rules the heavy quark masses are fixed by the two-point sum
rules for bottomonia and charmonia with the precision of 20 MeV. In our
consideration the quark masses are equal to $m_b$=4.6 GeV, $m_c$=1.4 GeV, and
we use $m_s$=0.15 GeV, which agrees with the various estimates \cite{ckm}. The
uncertainties in the values of form factors are basically determined by the
variation of $b$-quark mass, while changing the other quark masses in the
ranges $m_s=0.14\div 0.16$ GeV and $m_c =1.35\div 1.45$ GeV, results in the
uncertainty less than 2\%. In Figs. \ref{Faomom-fig}, \ref{fpmom-fig} and
\ref{3D-fig} we present the results in the scheme of  spectral density moments. 

\setlength{\unitlength}{1mm}
\begin{figure}[ph]
\begin{center}
\begin{picture}(100, 80)
\put(0, 0){\epsfxsize=11cm \epsfbox{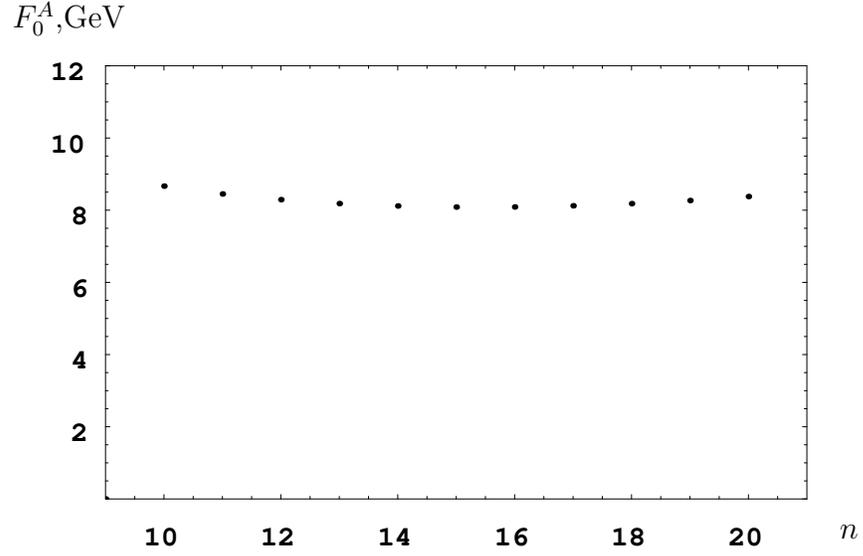}}
\put(110, 2){$n$}
\put(0, 70){$F^A_0$,GeV}
\end{picture}
\end{center}
\caption{ $F^A_0$ in the scheme of the spectral density moments; $n$ is the
number of moment with respect to the square of momentum in the $\bar b c$
channel. The number of moment with respect to the square of momentum in the
$\bar b s$ channel is equal to 1 at $E_c=1.2~$GeV.} 
\label{Faomom-fig}
\end{figure}

\setlength{\unitlength}{1mm}
\begin{figure}[ph]
\begin{center}
\begin{picture}(100, 80)
\put(0, 0){\epsfxsize=11cm \epsfbox{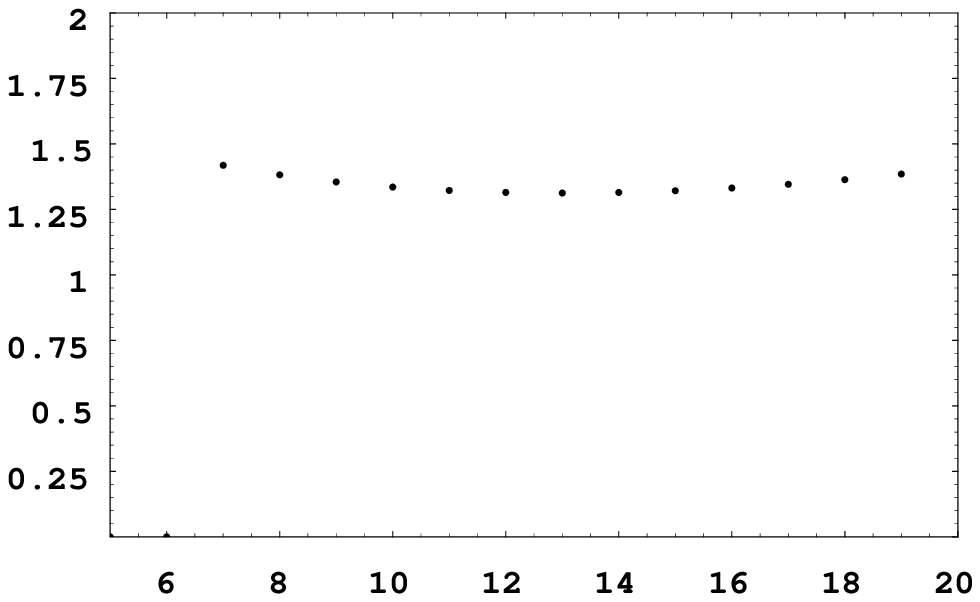}}
\put(115, 2){$n$}
\put(8, 70){$f_+$}
\end{picture}
\end{center}
\caption{ $f_+$ in the scheme of the spectral density moments; $n$ is the
number of moment with respect to the square of momentum in the $\bar b c$
channel. The number of moment with respect to the square of momentum in the
$\bar b s$ channel is equal to 1 at $E_c=1.2~$GeV.} 
\label{fpmom-fig}
\end{figure}

\setlength{\unitlength}{1mm}
\begin{figure}[th]
\begin{center}
\begin{picture}(100, 80)
\put(0, 0){\epsfxsize=11cm \epsfbox{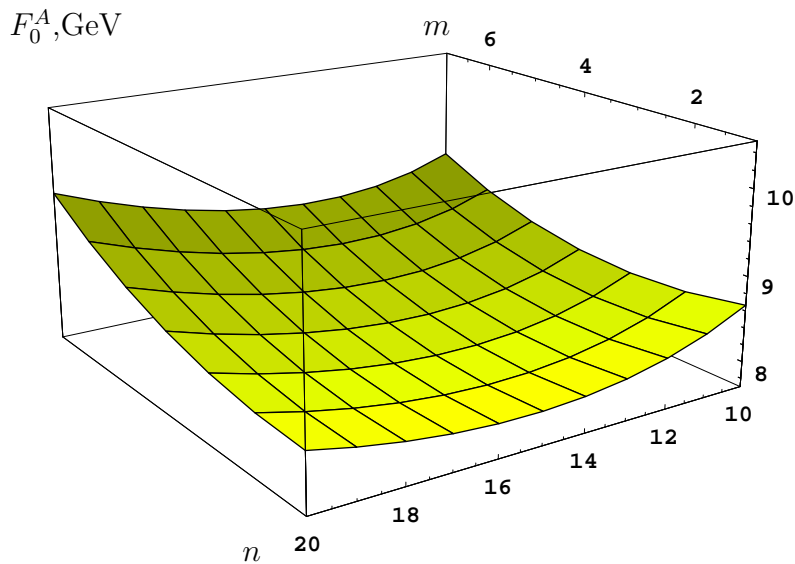}}
\put(31, 0){$n$}
\put(55,70){$m$}
\put(0, 70){$F^A_0$,GeV}
\end{picture}
\end{center}
\caption{ $F^A_0$ in the scheme of the spectral density moments; $n$ is the
number of moment with respect to the square of momentum in the $\bar b c$
channel, $m$ is the number of moment with respect to the square of momentum in
the $\bar b s$ channel at $E_c=1.2~$GeV.} 
\label{3D-fig}
\end{figure}

\begin{figure}[ph]
\begin{center}
\begin{picture}(100, 100)
\put(0, 0){\epsfxsize=11cm \epsfbox{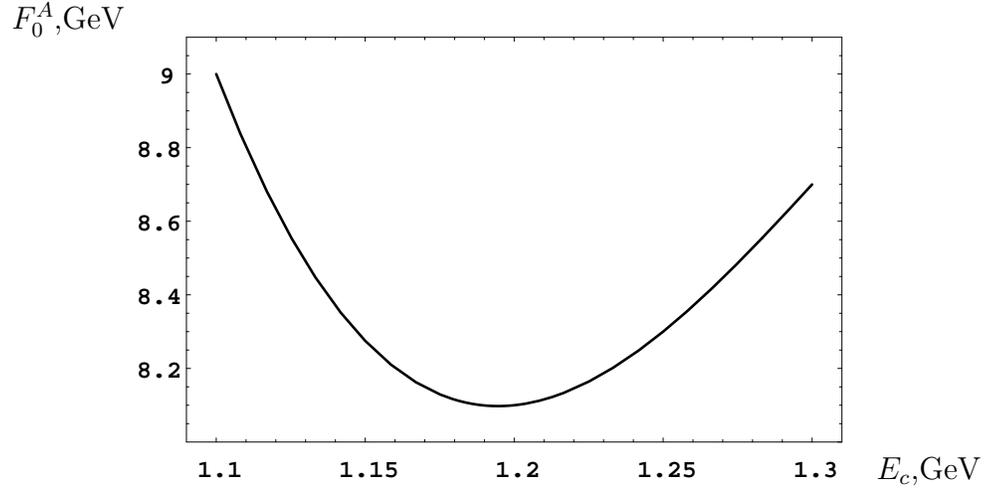}}
\put(100, 10){$E_c$,GeV}
\put(-15, 70){$F^A_0$,GeV}
\end{picture}
\end{center}
\caption{The dependence of $F^A_0$ on the $\bar b$s threshold energy $E_{c}$,
determining the region of resonance contribution.  } 
\label{FaoEc-fig}
\end{figure}

\begin{figure}[ph]
\begin{center}
\begin{picture}(100, 80)
\put(0, 0){\epsfxsize=11cm \epsfbox{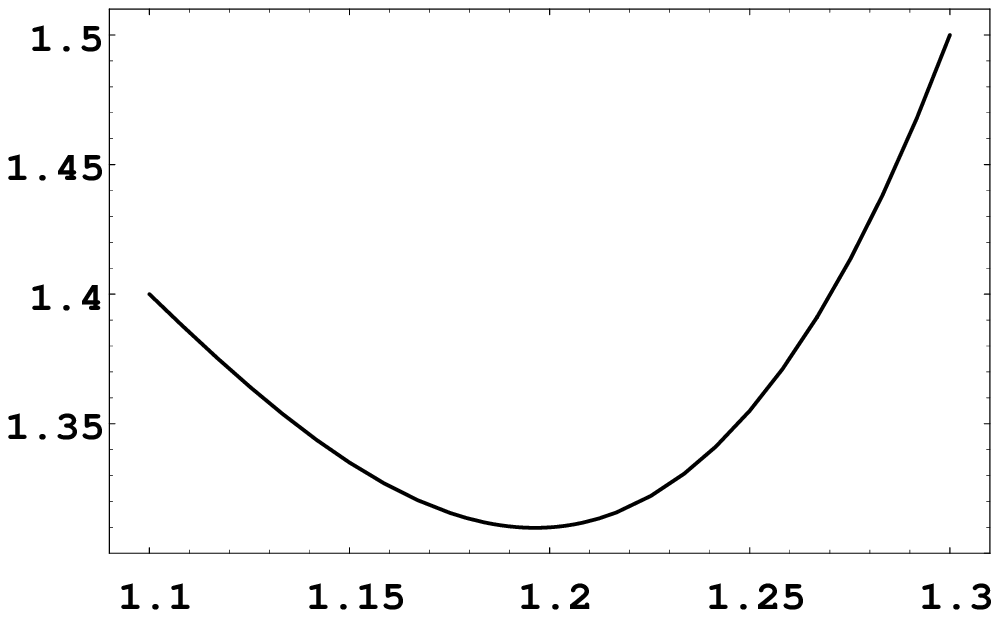}}
\put(94,3){$E_c$,GeV}
\put(0, 70){$f_+$}
\end{picture}
\end{center}
\caption{The dependence of $f_+$ on the $\bar b$s threshold energy $E_{c}$,
determining the region of resonance contribution.} 
\label{fpEc-fig}
\end{figure}

We have investigated the dependence of form factors on the $\bar b s$ threshold
energy of continuum in the range $E_c=1.1 \div 1.3$ GeV. The characteristic
forms of this dependence are shown in Figs. \ref{FaoEc-fig} and \ref{fpEc-fig}.
We see that the optimal choice for the $\bar b s$ system threshold energy is
1.2 GeV. In Table \ref{form} we present the results of sum rules for the form
factors in comparison with estimates in the framework of potential models
\cite{Bcstatus,Kis}. We see a good agreement of estimates in the QCD sum
rules with the values in the quark model. For the sake of completeness the
quark model expressions for the form factors are given in Appendix B.

\begin{table}[th]
\begin{center}
\begin{tabular}{|c|c|c|c|c|c|c|}
\hline
 Method & $f_+$ & $f_-$ & $F_{V}, \mbox{GeV}^{-1}$ & $F_{0}^{A}, \mbox{GeV}$ &
 $F_{+}^{A},
 \mbox{GeV}^{-1}$ 
 & $F_{-}^{A}, \mbox{GeV}^{-1}$ \\
\hline 
 This paper&1.3&-5.8&1.1&8.1&0.2&1.8 \\
\hline
Potential model \cite{Bcstatus}&1.1&-5.9&1.1&8.2&0.3&1.4 \\
\hline
\end{tabular}
\end{center}
\caption{The form factors of $B_{c}$ decay modes into the $B_{s}$ and
$B_{s}^{*}$ mesons at $q^{2}=0$.}
\label{form}
\end{table}

In \cite{Colangelo} the form factors were derived using the similar SR
technique but without the Coulomb-like corrections in the $\bar b c$ system,
which enhance the form factors about three times, as we have found.

Let us discuss the uncertainties in the sum rules and other approaches. So,
the potential models \cite{LusMas,CF} with similar choices of parameters result
in the form factor values, which are slightly model-dependent. The
corresponding accuracy is about 10\%. Then, we expect that the potential models
give good reference points for the appropriate numerical values. 

The accuracy of sum rules under consideration is basically determined by the
variation of heavy quark masses. Indeed, the significant $\alpha_s$ correction
to the leptonic constant of $B_s$ meson should cancel the same factor for the
renormalization of quark-meson vertex in the triangle diagram. The dependence
on the choice of threshold energy in the $\bar b s$-channel can be optimized
and, hence, minimized. The variation of threshold energy in the $\bar b
c$-channel give the error less than 1\%. The effective coulomb constant is
fixed from the two-point sum rules for the heavy quarkonium, and its variation
is less than 2\%, which gives the same uncertainty for the form factors. The
heavy quark masses are determined by the two-point sum rules for the heavy
quarkonia, too. However, their variations result in the most essential
uncertainty. Summing up all of mentioned variations we estimate $\delta f/f
\simeq 5$\%.

The enhancment of form factors in the decays of heavy quarkonium considered in
the framework of sum rules was also found in the decays $B_c\to \psi l\nu$
\cite{KT,Onish}. It is important to stress that the current consideration
removes the contradiction between the estimates in OPE, potential models and
the values in the SR. We have calculated the form factors in the SR consistent
with the values in the OPE and potential models.
 
We use the three-point sum rules to determine the dependence on $q^2$ of
form factors in vicinity of $q^2=0$, where the method works even in the
approximation by a bare quark loop \cite{SR3pt}. Naively, we expect a simple
pole form 
\begin{equation}
f(q^2)=\frac{f(0)}{1-\frac{q^2}{M^2_{pole}}},
\label{pole}
\end{equation}
so that the first derivative $f^\prime (0) = f(0)\frac{1}{M^2_{pole}}$ can be
evaluated in the framework of sum rules to estimate the ``pole'' mass
$M_{pole}$, which may deviate from the value of physical mass of corresponding
bound state. Of course, the bare quark loop approximation cannot be justified
at $q^2\sim m^2_{pole}$, while the three-point sum rules in the form of double
dispersion relation cannot be explored in the region of resonances at $q^2>0$
because of so-called ``non--Landau'' singularities indicating the presence of
strongly bound states in the $q^2$-channel. That is why we calculate the form
factors as well as their Borel transforms at $q^2<0$. 

Numerically, we have found $M_{pole}$=1.3$\div$1.4 GeV  for the form factors
with the $B^*_s$ decay modes, and $M_{pole}$=1.8$\div$1.9 GeV for the decay
form factors with the $B_s$ modes. 

The semileptonic widths are presented in Table \ref{slwidth}. We
have supposed the quark mixing matrix element $|V_{cs}|$=0.975 \cite{ckm}. The
mesons masses are equal to $M_{B_c}$=6.25 GeV, $M_{B_s}$=5.37 GeV,
$M_{B_s^*}$=5.41 GeV \cite{PDG}.  

\begin{table}[ph]
\begin{center}
\begin{tabular}{|c|c|c|}
\hline
 mode & $\Gamma$, $ 10^{-14}\ \mbox{GeV}$ & BR, $\%$ \\
\hline
 $B_{s}e^{+}\nu_{e}$ & 5.8 & 4.0 \\
\hline
 $B_{s}^{*}e^{+}\nu_{e}$ & 7.2 & 5.0 \\
\hline 
\end{tabular}
\end{center}
\caption{The widths of semileptonic $B_{c}$ decay modes and the
branching fractions calculated at $\tau_{B_{c}}=0.46$ ps.}
\label{slwidth}
\end{table}

These results agree with the values obtained in the framework of covariant
quark model \cite{Bcstatus}:
$\Gamma$($B_{s}e^{+}\nu_{e}$)=$4.7\cdot10^{-14}~$GeV,
$\Gamma$($B_{s}^*e^{+}\nu_{e}$)=$7.4\cdot10^{-14}~$GeV, as we could expect
looking at Table \ref{form}.

\section{The symmetry relations}
At the recoil momentum close to zero, the heavy quarks in both the initial and
final states have small relative velocities inside the hadrons, so that the
dynamics of heavy quarks is essentially nonrelativistic. This allows us to use
the combined NRQCD/HQET approximation in the study of mesonic form factors. The
expansion in the small relative velocities to the leading order leads to
various relations between the different form factors. Solving these relations
results in the introduction of an universal form factor (an analogue of the
Isgur-Wise function) at $q^2 \rightarrow q^2_{max}$.

We consider the soft limit
\begin{eqnarray}
v_1^{\mu} \neq v_2^{\mu}, \nonumber\\ 
w=v_1 \cdot v_2 \rightarrow 1,
\label{cs}
\end{eqnarray}
where $v_{1,2}^{\mu} = p_{1,2}^{\mu}/\sqrt{p_{1,2}^2}$ are the four-velocities
of heavy mesons in the initial and final states. The study of region
(\ref{cs}) is reasonable enough, because in the rest frame of $B_c~$meson 
($p_1^{\mu} = (\sqrt{p_1^2}, \vec 0)$), the four-velocities differ only by a
small value $|\vec p_2|$ $(p_2^{\mu} = (E_2, \vec p_2)$, whereas their
scalar product $w$ deviates from unity only due to a term, proportional to the
square of $|\vec p_2|$: 
$w = \sqrt{1 + \frac{|\vec p_2|^2}{p_2^2}}\sim 1 + \frac{1}{2}\frac{|\vec
p_2|^2}{p_2^2}$.
Thus, in the linear approximation at $|\vec p_2|\to 0$, relations
(\ref{cs}) are valid and take place.
Here we would like to note, that (\ref{cs}) generalizes the investigation of
\cite{Jenkins}, where the case of $v_1 = v_2$ was considered. This condition
severely restricts the relations of spin symmetry for the form factors and, as
a consequence, it provides a single connection between the form factors. In the
soft limit of zero recoil we find

\begin{equation}
 \tilde v_3^{\mu}=-\frac{1}{2}(v_1^{\mu}+v_2^{\mu})
\end{equation}
for the four-velocity of spectator $b$-quark, and
\begin{equation}
\tilde v_1^{\mu} = v_1^{\mu} + \frac{m_3}{2m_1}(v_1 - v_2)^{\mu}
\end{equation}
for the decaying $c$-quark. The matrix element of $J_{\mu} = \bar
Q_1\Gamma_{\mu} q_2$
with the spin structure $\Gamma_{\mu} = \{\gamma_{\mu}, \gamma_5\gamma_{\mu}\}$
has the form 
\begin{eqnarray}
\langle H_{Q_1\bar Q_3}|J_{\mu}|H_{q_2\bar Q_3}\rangle &=& 
tr[\Gamma_{\mu}(1 + \tilde v_1^{\mu}\gamma_{\mu})\Gamma_1
(1 + \tilde v_3^{\nu}\gamma_{\nu})\cdot \nonumber \\
&& \Gamma_2 \rho_{light}]\cdot h,
\label{29}
\end{eqnarray}
where $\Gamma_{1}$ determines the spin state in the heavy meson $Q_1\bar
Q_3$ (in our case it is pseudoscalar, so that $\Gamma_{1} = \gamma_5$),
$\Gamma_2$ determines the spin wave function of quarkonium in the final
state $\Gamma_2 = \{\gamma_5,  \epsilon^{\mu}\gamma_{\mu}\}$ for the
pseudoscalar and vector states, respectively  ($H = {P,V}$). The `propagator of
the light quark'\footnote{More strictly, we determine the spin structure of
matrix element and the term given by light degrees of freedom.} is taken in a
general form
\begin{equation}
 \rho_{light}=1+B(\slashchar v_2-\slashchar v_1)+C(\slashchar
 v_2+\slashchar v_1)+D\slashchar v_2 \slashchar
 v_1,
\label{Lprop} 
\end{equation}
where $B, C, D$ are the functions of $w$. The quantity $h$ in (\ref{29}) at
$w\to 1$ is an universal factor independent of the spin state of meson. So, for
the form factors, discussed in our paper, we have
\begin{eqnarray}
\langle P_{Q_1\bar Q_3}|\bar Q_1\gamma^{\mu} Q_3|P_{q_2\bar Q_3}\rangle &=& 
(c_1^{P}\cdot v_1^{\mu} + c_2^{P}\cdot v_2^{\mu})\cdot h, \\
\langle P_{Q_1\bar Q_3}|\bar Q_1\gamma^{\mu} Q_3|V_{q_2\bar Q_3}\rangle &=& 
i c_V\cdot\epsilon^{\mu\nu\alpha\beta}\epsilon_{\nu}v_{1\alpha}v_{2\beta}\cdot
h,
\\
\langle P_{Q_1\bar Q_3}|\bar Q_1\gamma_5\gamma^{\mu} Q_3|V_{q_2\bar Q_3}\rangle
&=& 
(c_{\epsilon}\cdot\epsilon^{\mu} + c_1\cdot v_1^{\mu}(\epsilon\cdot v_1) + 
c_2\cdot v_2^{\mu}(\epsilon\cdot v_1))\cdot h,
\end{eqnarray}
where
\begin{eqnarray}
c_{\epsilon} &=& -2,\nonumber\\
c_V &=& -1-\tilde B-\frac{m_3}{2m_1},\\
c_1^{P} &=& 1-\tilde B+\frac{m_3}{2m_1},\nonumber\\ 
c_2^{P} &=& 1+\tilde B-\frac{m_3}{2m_1},\nonumber 
\end{eqnarray}
and $\tilde B=\frac{B-2D}{1+C}$. The rest coefficients $c_{1,2}$ depend on the
C and D parameters. We have the symmetry relations for the following form
factors\footnote{To remove an error in \cite{Onish} the analogous second
relation for the $B_c \rightarrow J/\Psi(\eta_c)$ transition should have the
missed factor 2 in front of $c_{\epsilon}$.}:
\begin{eqnarray}
f_{+}(c_1^{P}\cdot{\cal M}_2 - c_2^{P}{\cal M}_1) - f_{-}(c_1^{P}\cdot{\cal
M}_2 + c_2^{P}\cdot{\cal M}_1) &=& 0,\nonumber\\
F_{0}^{A}\cdot c_V -2 c_{\epsilon}\cdot F_V{\cal M}_1{\cal M}_2 &=& 0,
\label{Fsym}\\
F_{0}^{A}c_1^{P} + c_{\epsilon}\cdot{\cal M}_1(f_{+} + f_{-}) &=& 0, \nonumber
\end{eqnarray}
where ${\cal M}_1=m_1+m_3$, ${\cal M}_2=m_2+m_3$. Equating the second relation
in (\ref{Fsym}), for example, we obtain
\begin{equation}
\tilde B=-\frac{2m_1+m_3}{2m_1}+\frac{4m_3(m_1+m_3)F_V}{F_0^A}\simeq 10.0,
\end{equation}
where all form factors are taken at $q^2_{max}$. Substituting $\tilde B$ in
first and third relations, we get $f_+\simeq 2.0$ and $f_-\simeq -8.3$. These
values have to be compared with the corresponding form factors obtained in the
QCD sum rules: $f_+(q^2_{max})=1.8$ and $f_-(q^2_{max})=-8.1$, where we suppose
the pole like behaviour of form factors (see Eq.(\ref{pole})). Thus, we find
that in the QCD sum rules, relations (\ref{Fsym}) are valid with the accuracy
better than $10\%$ at $q^2=q^2_{max}$. The deviation could increase at
$q^2<q^2_{max}$ because of variations in the pole masses governing the
evolution of form factors. However, in $B_c^+ \rightarrow B_s^{(*)}l^+\nu$
decays the phase space is restricted, so that the changes of form factors are
about 50\%, while their ratios develop more slowly. 

\section{Nonleptonic decays and the lifetime}

The hadronic decay widths can be obtained on the basis of assumption on the
factorization for the weak transition between the quarkonia and the final
two-body hadronic states. For the dominant nonleptonic decay modes
$B_c^+\rightarrow B_s^{(*)}\pi^+(\rho^{+})$ the effective Hamiltonian can be
written down as 
\begin{equation}
H_{eff}=\frac{G_F}{2 \sqrt{2}}V_{cs}V_{ud}^*\{C_+(\mu)O_{+}+C_-(\mu)O_-\},
\label{Heff}
\end{equation}
where 
\begin{equation}
O_{\pm}=(\bar u_i\gamma_{\nu}(1-\gamma_5)d_i)(\bar
s_j\gamma^{\nu}(1-\gamma_5)c_j) \pm (\bar u_i\gamma_{\nu}(1-\gamma_5)d_j)(\bar
s_i\gamma^{\nu}(1-\gamma_5)c_j),
\end{equation}
where $i,j$ run over the colors. 
 The factors $C_{\pm}(\mu)$ account for the strong corrections to the
 corresponding four-fermion operators caused by hard gluons. The review on the
 evaluation of $C_{\pm}(\mu)$ can be found in \cite{NLO}. In the present paper, 
 dealing with the QCD sum rules in the leading order over $\alpha_s$, we
 explore the $C_{\pm}(\mu)$-evolution to the leading log accuracy.  
 The $B_c^+\rightarrow B_s \pi^+$ amplitude, for example, takes the form
\begin{equation}
 A(B_c^+ \rightarrow B_s \pi^+)=\frac{G_F}{\sqrt{2}}V_{cs}V_{ud}a_1(\mu)\langle
 \pi^+|\bar u\gamma_{\nu}(1-\gamma_5)d|0 \rangle \langle
 B_s|\bar s\gamma^{\nu}(1-\gamma_5)c|B_c \rangle , 
\end{equation} 
where $a_1(\mu)= \frac{1}{2 N_{c}} (C_+(\mu)(N_c+1)+C_-(\mu)(N_c-1))$ at
$N_c=3$ being the number of colors. In our calculations we put the following
light meson parameters: $m_{\pi^+}$=0.14 GeV, $m_{\rho^+}$=0.77 GeV,
$f_{\pi^+}$=0.132 GeV, $f_{\rho^+}$=0.208 GeV. The results are collected in
Table \ref{width}. 

\begin{table}[th]
\begin{center}
\begin{tabular}{|c|c|c|}
\hline
 mode &  $\Gamma$, $ 10^{-14}\ \mbox{GeV}$ & BR, $\%$ \\
\hline
 $B_{s}\pi^{+}$ & 15.8 $a_{1}^{2}$ & 17.5 \\
\hline
$B_{s}\rho^{+}$ & 6.7 $a_{1}^{2}$ & 7.4 \\
\hline
 $B_{s}^{*}\pi^{+}$ & 6.2 $a_{1}^{2}$ & 6.9 \\
\hline
 $B_{s}^{*}\rho^{+}$ & 20.0 $ a_{1}^{2}$ & 22.2 \\
\hline 
\end{tabular}
\end{center}
\caption{The widths of dominant nonleptonic $B_{c}$ decay modes due to $c
\rightarrow s$ transition and the
branching fractions calculated at $\tau_{B_{c}}=0.46$ ps. We put
$a_{1}$=1.26.}
\label{width}
\end{table}

It is worth noting that the sum of widths for transitions $B_c^+ \rightarrow
B_s(B_s^*)\pi^+(\rho^+)$ is $10\%$ larger than the width for the transition
$B_c^+ \rightarrow B_s(B_s^*)+light~hadrons$, which is calculated using the
simple formula
$$
\Gamma[B_c^+ \rightarrow
B_s(B_s^*)+light~hadrons]=N_c~a_1^2(\mu)~\Gamma[B_c^+
\rightarrow B_s(B_s^*)e^+\nu_e],
$$ 
where we neglect the contributions given by the modes with the factor $a_2$
instead of $a_1$. In addition, the deviation between these estimates can be
caused by the corresponding `bag' factor appearing in the formulation of
factorization approach and vacuum saturation in the connection of leptonic form
factors to the hadronic ones. The modern lattice estimates show that the `bag'
parameters are about 7\% less than 1 \cite{Lellouch}.

In the parton approximation we could expect 
$$ 
\Gamma[B_c^+ \rightarrow B_s^{(*)}+light~hadrons] = (2C_+^2(\mu)+C_-^2(\mu))
\Gamma[B_c^+ \rightarrow B_s^{(*)}e^+\nu_e], 
$$
which results in the estimate very close to the value obtained as the sum of
exclusive modes at $\mu>0.9~$GeV. The deviation between these two estimates
slightly increase at $\frac{m_c}{2}<\mu<0.9~$GeV. Concerning the comparison of
hadronic width summing up the exclusive decay modes with the estimate based on
the quark-hadron duality, we insist that the deviation between these two
estimates is less that 10\% and, hence, it cannot be treated as an essential
argument against the validity of our calculations.

We estimate the lifetime using the fact that the dominant modes of the
$B_c~$meson decays are the $c \rightarrow s,~b \rightarrow c~$ transitions with
the $B_s^{(*)}$ and J/$\psi$, $\eta_c$ final states respectively, and the
electroweak annihilation \footnote{The $\bar b \rightarrow \bar c c \bar s$
transition is negligibly small in the $B_c$ decays because of destructive Pauli
interference for the charmed quark in the initial state and the product of
decay \cite{Buchalla}.}.

We stress that in the $B_c\to B_s$ decays caused by the weak decays of charmed
quark the possible hadronic final states are the charged mesons $\pi$, $\rho$
and $K$ or the multi-particle states like $\pi\pi$,  $\pi\pi\pi$ or $K\pi$.
First, the states with the kaon are suppressed by the Cabibbo angle, and we
neglect their contributions in the total nonleptonic width of $B_c$ (the
corresponding error of estimates is about 4\%). The method for the calculation
of multi-particle branching fractions was offered by Bjorken in his pioneering
paper devoted to the decays of hadrons containing heavy quarks \cite{Bjorken}.
He supposed a simple relation for the yields of pions, as given by the Poisson
distribution with the average value of pions determined by the energy release.
The Bjorken's model of multi-particle yields in the decays does not distinguish
the resonant and continuum final states as well as $B_s$ and $B_s^*$. So, we
check that the ratio of $R_{2\pi} = \Gamma(B_c^+\to B_s^{(*)}\pi^+)/
\Gamma(B_c^+\to B_s^{(*)}\rho^+)\approx 0.82$ calculated in the framework of
sum rules is close to the estimate $R_{2\pi} = \Gamma(B_c^+\to B_s^{(*)}\pi^+)/
\Gamma(B_c^+\to B_s^{(*)}\pi^+\pi^0)\approx 0.85$ given by Bjorken. Then, we
see that the non-resonant multi-particle states are suppressed in comparison
with the resonance yields. The same fact can be found, once we consider the
$K\pi$, $K\pi\pi$ and $K^*\pi$, $K\rho$ branching fractions in the decays of
$D$ mesons as measured experimentally. This consideration confirms us that we
take into account all of significant nonleptonic decay modes of $B_c$. In order
to estimate the contribution of non-resonant $3\pi$ modes of $B_c$ decays into
$B_s^{(*)}$ we use the Bjorken's technique, i.e. the Poisson distribution with
the average value corrected to agree with the non-resonant $3\pi$-modes in the
decays of $D$ mesons. We have chosen the following branching ratios: ${\rm
BR}(D^+\to K^-\pi^+\pi^+)=9.0\pm 0.6$\% and ${\rm BR}(D^+\to
K^-\pi^+\pi^+\pi^0|_{\rm non-resonant})=1.2\pm 0.6$\%. So, we have found $\bar
n\approx \frac{1}{8}$, which means that ${\rm BR}(B_c^+\to
B_s^{(*)}(3\pi)^+)\approx 0.2$\%, while ${\rm BR}(B_c^+\to
B_s^{(*)}(2\pi)^+|_{\rm non-resonant})\approx 3$\%. We see that the neglected
modes contribute to the total width of $B_c$ as a small fraction in the limits
of uncertainty envolved.

\setlength{\unitlength}{1mm}
\begin{figure}[th]
\begin{center}
\begin{picture}(100, 80)
\put(0, 0){\epsfxsize=11cm \epsfbox{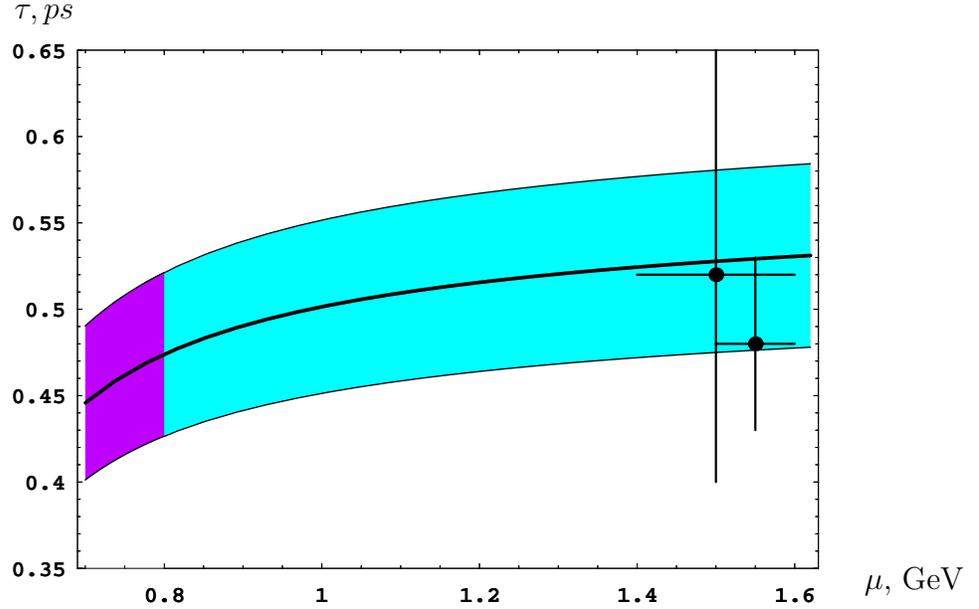}}
\put(115,3){$\mu$, GeV}
\put(2, 79){$\tau, ps$}
\end{picture}
\end{center}
\caption{The dependence of $B_c$ meson lifetime on the scale $\mu$ in the
effective Hamiltonian (\ref{Heff}). The shaded region shows the uncertainty of
estimates, the dark shaded region is the preferable choice as given by the
lifetimes of charmed mesons. The dots represent the values in the OPE
approach.} 
\label{Tau-fig}
\end{figure} 

The width of beauty decay in the sum rules was derived using the similar
methods in \cite{Onish}: $\Gamma(B^+_c \rightarrow \bar c
c+X)=(28\pm5)\cdot10^{-14}~$GeV. The width of the electroweak annihilation is
taken from \cite{Bcstatus} as $12\cdot10^{-14}~$GeV. 

In Fig. \ref{Tau-fig} we present the $B_c~$meson lifetime calculated in the QCD
SR under consideration. We also show the results of the lifetime evaluation in
the framework of Operator Product Expansion in NRQCD \cite{Buchalla,BcOnish}. 

In contrast to OPE, where the basic uncertainty is given by the variation of
heavy quark masses, these parameters are fixed by the two-point sum rules for
bottomonia and charmonia, so that the accuracy of SR calculations for the total
width of $B_c$ is determined by the choice of scale $\mu$ for the hadronic weak
lagrangian in decays of charmed quark. We show this dependence in Fig.
\ref{Tau-fig}, where $\frac{m_c}{2} < \mu < m_c$ and the dark shaded region
corresponds to the scales preferred by data on the charmed meson lifetimes. The
discussion on the optimal choice of scale in hadronic decays is addressed in
the next section.

\section{Discussion on the lifetimes of heavy hadrons}

At present the ordinary prescription for the normalization point of lagrangian
generating the nonleptonic decays of heavy quark $Q$, is $\mu \simeq m_Q$. The
motivation is the following: the characteristic scale is determined by the
energy release given by the heavy quark mass. Therefore, we can argue the
operator product expansion in the inverse powers of $m_Q$, wherein we can
factorize the Wilson coefficients taken in the perturbative QCD and the matrix
elements of operators over the hadronic states with $\mu$ usually posed to
$m_Q$. This prescription is in a qualitative agreement with the
current data on the measured lifetimes of charmed and beauty hadrons and their
branching ratios for the semileptonic decay modes, say.

Let us consider this issue in more details. To the moment, the analysis of
decays in the QCD sum rules is restricted by the leading order (LO) in
$\alpha_s$ (except the Coulomb-like corrections in the heavy quarkonia). The
corresponding parameters, the heavy quark masses, are also fixed to the same
order (they have to be reevaluated in next-to-leading order). Therefore, for
the sake of consistency, we use the LO expressions, which are given by the
partonic approximation improved by $1/m_Q$-corrections in HQET or NRQCD. In
this way, we can write down the following formulae:

1. The semileptonic branching fraction of $D^0$ meson is given by
\begin{equation}
{\rm BR}_{sl}[D^0] = \frac{1}{2+2 C_+^2(\mu)+C_-^2(\mu)}.
\end{equation}

2. The difference between the total widths of charmed mesons is determined by
the Pauli interference in decays of $D^+$, so that \cite{Voloshinnew}
\begin{equation}
\Gamma[D^+]-\Gamma[D^0] = \cos^4\theta_c G_{\rm F}^2 \frac{m_c^3
f_D^2}{8\pi}\left[ (C_+^2(\mu)-C_-^2(\mu)) B + \frac{1}{3}
(C_+^2(\mu)+C_-^2(\mu)) \tilde B \right],
\end{equation}
where $\theta_c$ is the Cabibbo angle, $f_D$ is the leptonic constant of
$D$ meson, and the `bag' constants are defined by
\begin{eqnarray}
\langle D| (\bar c \Gamma_\mu q) (\bar q \Gamma_\mu c) |D\rangle & = & ~f_D^2
M_D^2 B, \\
\langle D| (\bar c \Gamma_\mu c) (\bar q \Gamma_\mu q) |D\rangle & = &
\frac{1}{3} f_D^2 M_D^2 \tilde B, 
\end{eqnarray}
with $\Gamma_\mu = \gamma_\mu (1-\gamma_5)$.

The similar expression can be derived for the beauty mesons \cite{Voloshinnew}
\begin{equation}
\Gamma[B^+]-\Gamma[B^0] = |V_{bc}|^2 G_{\rm F}^2 \frac{m_b^3
f_B^2}{8\pi}\left[ (C_+^2(\mu)-C_-^2(\mu)) B + \frac{1}{3}
(C_+^2(\mu)+C_-^2(\mu)) \tilde B \right].
\end{equation}

3. The semileptonic branching fraction of $B^0$ meson is given by
\begin{equation}
{\rm BR}_{sl}[B^0] = \frac{1}{2+0.22+[2 C_+^2(\mu)+C_-^2(\mu)](1+k)},
\end{equation}
where the fraction of $0.22$ is due to the $\tau\nu_\tau$-contribution, while
the value of $k$ denotes the fraction of $b\to c\bar c s$ transition in the
nonleptonic decays. In the same way, the average yield of charm in the decays
of beauty mesons is equal to
\begin{equation}
n_c = \frac{2+0.22+[2 C_+^2(\mu)+C_-^2(\mu)](1+2 k)}{2+0.22+[2
C_+^2(\mu)+C_-^2(\mu)](1+k)}.
\end{equation}
As for numerical applications, one usually puts

$\mu_D =m_c$ in decays of $D$ mesons,

$\mu_B =m_b$ in decays of $B$ mesons,

$f_D\approx f_B \approx 200$ MeV, and

$B=\tilde B=1$ naively motivated by nonrelativistic potential models. 

At $k=0.4$ \cite{MNeu}, this set (marked as SETMQ column) results in the
estimates shown in Table \ref{mu} in comparison with the experimental data
\cite{PDG}.

\begin{table}[th]
\begin{center}
\begin{tabular}{|l|c|c|c|c|}
\hline
quantity & exp. & SETMQ & SETMQ$_2$ & SETH \\[1mm]
\hline 
${\rm BR}_{sl}[D^0]$, \%             & $8.1\pm 1.1$    & 15.4 & 15.4 &
8.6\\[1mm]
$\Gamma[D^+]-\Gamma[D^0]$, ps$^{-1}$ & $-1.56\pm 0.03$ & -1.26 & -0.19 & -1.53
\\[1mm]
${\Gamma[D^0]}/{\Gamma[D_s]}$    & $1.12\pm 0.05$  & 1.00 & 1.00 & 1.11\\
[1mm]
${\rm BR}_{sl}[B^0]$, \%             & $10.45\pm 0.21$ & 14 & 14 & 10.2\\ 
[1mm]
$\Gamma[B^+]-\Gamma[B^0]$, ps$^{-1}$ & $-0.043\pm 0.017$ & -0.022 & 0.024 &
-0.044\\[1mm]
${\Gamma[B^0]}/{\Gamma[B_s]}$    & $1.00\pm 0.05$  & 1.00 & 1.00 &
1.05\\[1mm]
$n_c$                                & $1.12\pm 0.05$ & 1.20 & 1.20 &
1.12\\[1mm]
${\Gamma[B^0]}/{\Gamma[\Lambda_b]}$ 
                                     & $0.81\pm 0.05$ & 1.00 & 1.00 & 0.81\\
\hline
\end{tabular}
\end{center}
\caption{The comparison of theoretical estimates at various sets of
parameters with the experimental data on the decays of heavy mesons.}
\label{mu}
\end{table}

First, we note that the semileptonic width of $D^0$ is well described to the
given order and at chosen value of $m_c$, while its branching ratio is in a
valuable contradiction with the data indicating a more higher enforcement of
nonleptonic modes. Second, the qualitative agreement of predictions with the
measured differences of $\Gamma[D^+]-\Gamma[D^0]$ and $\Gamma[B^+]-\Gamma[B^0]$
is mainly based on the assumption of $\tilde B \approx 1$. Recent consideration
of charmed baryon lifetimes by M.Voloshin \cite{MVol} clearly drawn a
conclusion that the naive picture of color structure as given by the potential
models (i.e. the purely antisymmetric color-composition of valence flavors) is
significantly broken. A similar statement was obtained in the description of
$D^*$ meson production at HERA, where the authors of \cite{bkl} found that the
$O(\alpha_{em}\alpha_s^3)$-calculations for the differential cross sections of
$c\bar q$-pair composing the meson, are able to reproduce the measured spectra,
if we introduce the valuable contribution by the color-octet state in
addition to the singlet one. So, the four-quark singlet-operator results in
\begin{equation}
O_{(1)} = \langle D^*| (\bar c \gamma_\mu q) (\bar q \gamma_\mu c) |D^*\rangle
\cdot \left(-g_{\mu\nu}+\frac{p_\mu p_\nu}{p^2}\right) \frac{1}{12M},
\end{equation}
which is reduced to $O_{(1)} = |\Psi_{(1)}(0)|^2$ in the framework of
nonrelativistic potential model, where
\begin{equation}
|D^*\rangle = \sqrt{2M} \int \frac{d^3 q}{(2\pi)^3} \Psi_{(1)}(q)
\frac{\delta_{ij}}{\sqrt{3}} \cdot \frac{1}{\sqrt{2}} \bar c_i
\slashchar{\epsilon} q_j |0\rangle,
\end{equation}
with $\epsilon_\alpha$ denoting the polarization vector and $\Psi_{(1)}(q)$
being the wave function. 

The term of color-octet was parameterized by
\begin{equation}
O_{(8)} = \langle D^*| (\bar c \lambda^a\gamma_\mu q) (\bar q \lambda^a
\gamma_\mu c) |D^*\rangle \cdot \left(-g_{\mu\nu}+\frac{p_\mu
p_\nu}{p^2}\right) \frac{1}{64M},
\end{equation}
which is, in a similar way, can be represented as $O_{(8)} =
|\Psi_{(8)}(0)|^2$, if we introduce the additional Fock state
\begin{equation}
\sqrt{2M} \int \frac{d^3 q}{(2\pi)^3} \Psi_{(8)}(q)
\frac{\lambda^a_{ij}}{\sqrt{2}} n_a \cdot \frac{1}{\sqrt{2}} \bar c_i
\slashchar{\epsilon} q_j |0\rangle,
\end{equation}
where the $\bar c$ and $q$ fields represent the rapid valence quarks, and $n_a$
is a random color-vector determined by soft degrees of freedom inside the meson
(i.e. by the quark-gluon sea). We have $\langle n_a n_b\rangle = \delta_{ab}$
in the production, while $\langle n_a n_b\rangle = \frac{1}{8}\delta_{ab}$ in
decays.

The data on the $D^*$ production in DIS give $O_{(8)}/O_{(1)} \simeq
1.3$. We can analogously expect that in the $D$ meson the ratio of four-quark
matrix elements is close to that in $D^*$. So, $O_{(8)}[D]/O_{(1)}[D] \simeq
1$, where $O_{(1,8)}[D]$ can be obtained from the above expressions for $D^*$
by the substitution of $\gamma_5\gamma_\mu$ for $\gamma_\mu$ and removing the
transverse projector. Then we straightforwardly find
\begin{equation}
\tilde B = 1 + \frac{O_{(8)}[D]}{O_{(1)}[D]} \approx 2.
\end{equation}
Putting $\tilde B = 2$ into the SETMQ we get the values given in the SETMQ$_2$
column in Table \ref{mu}. So, the choice of $\mu=m_Q$ is in a deep
contradiction with the observed differences of total widths for the heavy
mesons at the most reasonable value of $\tilde B = 2$.

In this position we argue the following: There are other physical scales in the
problem, which are characteristic for two hadronic systems in the decay
process. The first system is the decaying hadron. The second is the transition
current $c\to s$ or $\bar b\to \bar c$, where the form factor behaviour versus
the transferred momentum is determined by the $c\bar s$ and $\bar b c$
states. Those hadronic systems have the following scales, being the average
squares of heavy quark momentum $\langle p^2\rangle$, which are
phenomenologically equal to
\begin{equation}
\mu^2_{c\bar u} = \mu^2_{b\bar d} = 2 T \bar \Lambda,
\end{equation}
where according to the potential models $\bar \Lambda \simeq 0.4$ GeV is the
binding energy of heavy quark (i.e. the constituent mass of light quark),
$T\simeq 0.45$ GeV is the average kinetic energy in the system. $T$ determines
the 2S-1S splitting and it is approximately independent of quark flavors.
Analogously, we put
\begin{equation}
\mu^2_{c\bar s} = \mu^2_{b\bar s} = 2 T \bar \Lambda_s,
\end{equation}
where $\bar \Lambda_s=\bar \Lambda+(m_{D_s}-m_D)=\bar
\Lambda+(m_{B_s}-m_B)\simeq 0.5$ GeV. For the $b\to c$ current we put
\begin{equation}
\mu^2_{c\bar b} = 2 T m_{bc},
\end{equation}
with $m_{bc} \simeq  m_b m_c/(m_b+m_c) \simeq  m_B m_D/(m_B+m_D)\approx 1.3$
GeV.

Let us suppose that the decay scale is given by the following combinations:
\begin{eqnarray}
\mu^2_{D}~ &=& \mu_{c\bar u}\cdot \mu_{c\bar s},\nonumber \\
\mu^2_{D_s} &=& \mu^2_{c\bar s}, \\
\mu^2_{B}~ &=& \mu_{b\bar u} \cdot\mu_{c\bar b},\nonumber \\
\mu^2_{B_s} &=& \mu_{b\bar s}\cdot \mu_{c\bar b},\nonumber 
\end{eqnarray}

At $\tilde B=2$, this set of parameters with $f_B\cong f_D\cong 175$ MeV, and
$k=0.18$ is represented by the SETH column in Table \ref{mu}. We see a good
agreement with the data. The result of $\mu_B$ and $k$ variations is also shown
in Fig. \ref{brnc}. 
\setlength{\unitlength}{1mm}
\begin{figure}[th]
\begin{center}
\begin{picture}(100, 80)
\put(0, 0){\epsfxsize=11cm \epsfbox{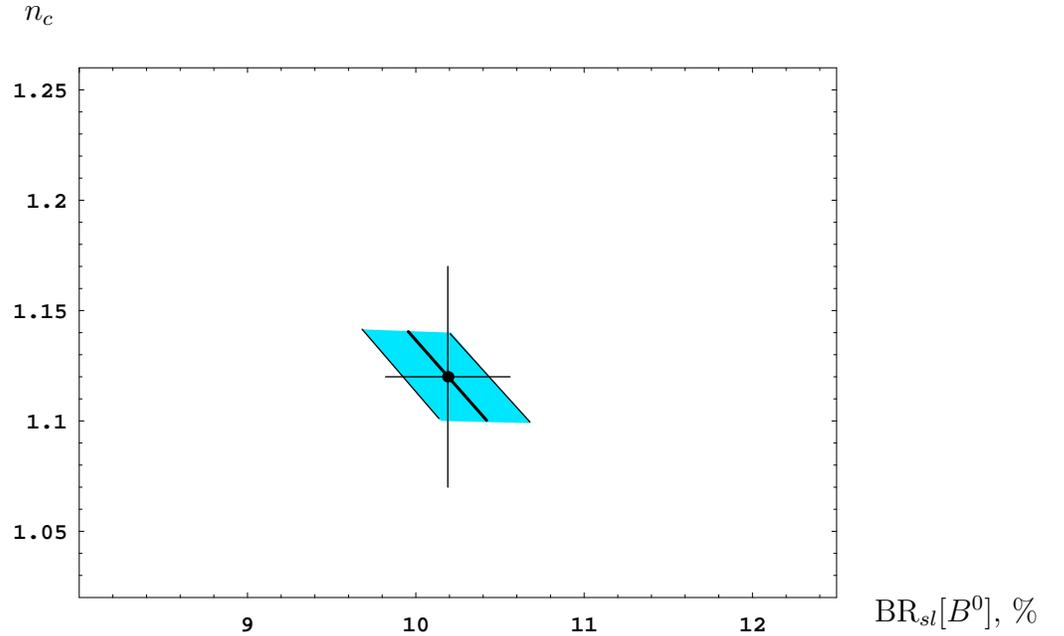}}
\put(115,3){BR$_{sl}[B^0]$, \%}
\put(2, 83){$n_c$}
\end{picture}
\end{center}
\caption{The predictions on the semileptonic branching ratio of $B^0$ meson and
the average yield of charmed quarks in its decays with the $\pm 3\%$-variation
of hadronic scale in the nonleptonic effective lagrangian and the change of
doubly charmed mode fraction $k=0.15\div 0.22$ in comparison with the ARGUS and
CLEO data shown by the dot with the error bars.} 
\label{brnc}
\end{figure} 

We stress that the theoretical OPE prediction for the
contribution of $b\to c\bar c s$ mode, $k=0.4$, is significantly overestimated,
to our opinion. Indeed, looking at the semileptonic decay with
$\tau^+\nu_{\tau}$ in the final state we see that at the same values of form
factors as in modes with light leptons the heavy lepton mode is suppressed
because of the restricted phase space, so that the reduction factor equals
0.22. The same effect has to take place in the $b$-quark decays with two
charmed quarks in the final state. So, since the sum of $D$ and $K$ masses is
greater than the $\tau$ mass we could expect even the greater suppression than
for the heavy lepton mode. Nevertheless, OPE operates with the quark masses at
so moderate release of energy, that certainly leads to the overestimation of
phase space in the decay of $b\to c\bar c s$. We use more realistic values for
$k$ close to $0.2$.

Next, the contribution of four-quark operators are suppressed in decays of
$\Lambda_b$ baryon. So, one expects that the deviation of its total width from
that of $B$ mesons is about 2-3\% \cite{IBig}. To the moment, the experimental
result is far away from this expectation (see Table \ref{mu}). We can reproduce
the data on the $\Lambda_b$ lifetime, if we introduce the light diquark scale
\begin{equation}
\mu^2_{ud} =  2 \frac{T}{2} \bar \Lambda,
\end{equation}
where $T/2$ corresponds to the half tension of color string inside the diquark.
Then we pose
\begin{equation}
\mu^2_{\Lambda_b} = \mu_{ud} \cdot\mu_{c\bar b},
\end{equation}
which result in a good agreement with the experimental data.

To finish this discussion we draw the conclusion: a probable way to reach the
agreement between the theoretical predictions and available experimental data
on the lifetimes and inclusive decay widths of heavy hadrons is to suggest the
different normalization points in the effective nonleptonic lagrangian for the
heavy quark weak decays as dependent on the hadron. This assumption provide us
with quite acceptable results to the leading order in $\alpha_s$. The variation
of normalization point shows the sensitivity of calculations to the higher
orders in the QCD coupling constant, which indicates, first, the necessity to
proceed with the higher orders, and, second, the appropriate choice of scale
can allow us to decrease the scale-dependent higher orders terms.

Thus, we suppose that the preferable choice of scale in the $c\to s$ decays of
$B_c$ is equal to
\begin{equation}
\mu^2_{B_c} = \mu_{c\bar b} \cdot\mu_{c\bar s}\approx (0.85\; {\rm GeV})^2,
\end{equation}
and at $a_1(\mu_{B_c}) =1.20$ in the charmed quark decays we predict
\begin{equation}
\tau[B_c] = 0.48\pm 0.05\;{\rm ps.}
\end{equation}

\section{Conclusion}
We have investigated the semileptonic decays of $B_c$ meson due to the weak
decays of charmed quark in the framework of three-point sum rules in QCD. We
have found the important role played by the Coulomb-like
${\alpha_s}/{v}$-corrections. As in the case of two-point sum rules, the
form factors are about three times enhanced due to the Coulomb renormalization
of quark-meson vertex for the heavy quarkonium $B_c$. We have studied the
dependence of form factors on the threshold energy, which determines the
continuum region of $\bar b s$ system. The obtained dependence has the
stability region, serving as the test of convergency for the sum rule method.
The HQET two-point sum rules for the leptonic constant $f_{B_s}$ and
$f_{B_s^*}$ have been reanalyzed to introduce the term caused by the product
of quark and gluon condensates. This contribution essentially improves the
stability of SR results for the leptonic constants of B mesons, yielding:
$f_B=140\div 170$ MeV.

We have studied the soft limit for the form factors in combined HQET/NRQCD
technique at the recoil momentum close to zero, which allows us to derive the
generalized relations due to the spin symmetry of effective lagrangian. The
relations  are in a good agreement with the full QCD results, which means that
the corrections to the form factors in both relative velocity of heavy quarks
inside the $\bar b c$ quarkonium and the inverse heavy quark masses are small
within the accuracy of the method. 
 
Next, we have studied the nonleptonic decays, using the assumption on the
factorization of the weak transition. The results on the widths and branching
fractions for various decay modes of $B_c$ are collected in Tables.

Finally, we have estimated the $B_c$ meson lifetime, and showed the dependence
on the scale for the hadronic weak lagrangian in decays of charmed quark
$$
\tau[B_c] = 0.48\pm 0.05\;{\rm ps.}
$$
Our estimates are in a good agreement with the theoretical predictions for the
lifetime in both the potential models and OPE as well as with the experimental
data.   

This work was in part supported by the Russian Foundation for Basic Research,
grants 99-02-16558 and 00-15-96645.

\section{Appendix A}
 For the perturbative spectral densities $\rho_i(s_1, s_2, Q^2)$  we have the
 following expressions \cite{Onish}:
\begin{eqnarray}
\rho_{+}(s_1, s_2, Q^2) &=& \frac{3}{2k^{3/2}}\{\frac{k}{2}(\Delta_1 +
\Delta_2) -
k[m_3(m_3 - m_1) + m_3(m_3 - m_2)] - \nonumber\\
&& [2(s_2\Delta_1 + s_1\Delta_2) - u(\Delta_1 + \Delta_2)]\\
&& \cdot [m_3^2 - \frac{u}{2} + m_1m_2 - m_2m_3 - m_1m_3]\}, \nonumber \\
\rho_{-}(s_1, s_2, Q^2) &=& - \frac{3}{2k^{3/2}}\{\frac{k}{2}(\Delta_1 -
\Delta_2) -
k[m_3(m_3 - m_1) - m_3(m_3 - m_2)] + \nonumber\\
&& [2(s_2\Delta_1 - s_1\Delta_2) + u(\Delta_1 - \Delta_2)]\\
&& \cdot [m_3^2 - \frac{u}{2} + m_1m_2 - m_2m_3 - m_1m_3]\}, \nonumber \\
\rho_{V}(s_1, s_2, Q^2) &=& \frac{3}{k^{3/2}}\{(2s_1\Delta_2 - u\Delta_1)(m_3 -
m_2)
\nonumber \\
&& + (2s_2\Delta_1 - u\Delta_2)(m_3 - m_1) + m_3k\}, \\
\rho_{0}^A(s_1, s_2, Q^2) &=& \frac{3}{k^{1/2}}\{
(m_1 - m_3)[m_3^2 + \frac{1}{k}(s_1\Delta_2^2 + s_2\Delta_1^2 -
u\Delta_1\Delta_2)]
\nonumber \\
&& - m_2(m_3^2 - \frac{\Delta_1}{2}) - m_1(m_3^2 - \frac{\Delta_2}{2}) \\
&& + m_3[ m_3^2 - \frac{1}{2}(\Delta_1 + \Delta_2 - u) + m_1m_2]\}, \nonumber\\
\rho_{+}^A(s_1, s_2, Q^2) &=& \frac{3}{2k^{3/2}}\{m_1[2s_2\Delta_1 - u\Delta_2
+
4\Delta_1\Delta_2 + 2\Delta_2^2]\nonumber \\
&& m_1m_3^2[4s_2 - 2u] + m_2[2s_1\Delta_2 - u\Delta_1] - m_3[2(3s_2\Delta_1 +
s_1\Delta_2)
\nonumber \\
&& - u(3\Delta_2 + \Delta_1) + k + 4\Delta_1\Delta_2 + 2\Delta_2^2 + m_3^2(4s_2
- 2u)] \\
&& + \frac{6}{k}(m_1 - m_3)[4s_1s_2\Delta_1\Delta_2 - u(2s_2\Delta_1\Delta_2 + 
s_1\Delta_2^2 + s_2\Delta_1^2)\nonumber \\
&& + 2s_2(s_1\Delta_2^2 + s_2\Delta_1^2)]\}, \nonumber \\
\rho_{-}^A(s_1, s_2, Q^2) &=& -\frac{3}{2k^{5/2}}\{
kum_3(2m_1m_3 - 2m_3^2 + u) + 12(m_1 - m_3)s_2^2\Delta_1^2 + \nonumber \\
&& k\Delta_2[(m_1 + m_3)u - 2s_1(m_2 - m_3)] + 2\Delta_2^2(k + 3us_1)(m_1-m_3) 
\nonumber \\
&& + \Delta_1[ku(m_2 - m_3) + 2\Delta_2(k - 3u^2)(m_1 - m_3)] + \\
&& 2s_2(m_1 - m_3)[2km_3^2 - k\Delta_1 + 3u\Delta_1^2 - 6u\Delta_1\Delta_2] - 
\nonumber\\
&& 2s_1s_2(km_3 - 3\Delta_2^2(m_1 - m_3))],\nonumber \\
\rho_{+}^{'A}(s_1, s_2, Q^2) &=& -\frac{3}{2k^{5/2}}\{ -2(m_1 - m_3)
[(k - 3us_2)\Delta_1^2 + 6s_1^2\Delta_2^2] + \nonumber \\
&& ku(m_1 - m_3)(2m_3^2 + \Delta_2) + ku^2m_3 + \Delta_1[ku(2m_1 - m_2 -
3m_3)\nonumber \\ 
&& - 2(m_1 - m_3)(ks_2 - k\Delta_2 + 3u^2\Delta_2)] - \\
&& 2s_1[(m_1-m_3)(2km_3^2 - 6u\Delta_1\Delta_2 - 3u\Delta_2^2) + \nonumber \\
&& 2s_2(km_3 + 3m_1\Delta_1^2 - 3m_3\Delta_1^2) + k\Delta_2(2m_1 - m_2 -
3m_3)]\},\nonumber \\
\rho_{-}^{'A}(s_1, s_2, Q^2) &=& \frac{3}{2k^{5/2}}\{ 2(m_1 - m_3)
[(k + 3us_2)\Delta_1^2 + 6s_1^2\Delta_2^2] + \nonumber \\
&& ku(m_1 - m_3)(2m_3^2 + \Delta_2) + ku^2m_3 + \Delta_1[ku(- 2m_1 - m_2 +
m_3)\nonumber \\ 
&& - 2(m_1 - m_3)(ks_2 - k\Delta_2 + 3u^2\Delta_2)] + \\
&& 2s_1[(m_1-m_3)(2km_3^2 - 6u\Delta_1\Delta_2 + 3u\Delta_2^2) - \nonumber \\
&& 2s_2(km_3 - 3m_1\Delta_1^2 + 3m_3\Delta_1^2) + k\Delta_2(2m_1 + m_2 -
m_3)]\}.\nonumber
\end{eqnarray}

Here $k = (s_1 + s_2 + Q^2)^2 - 4s_1s_2$, $u = s_1 + s_2 + Q^2$, $\Delta_1 =
s_1 - m_1^2 + m_3^2$ and $\Delta_2 = s_2 - m_2^2 + m_3^2$. $m_1, m_2$ and $m_3$
are the masses of quark flavours relevant to the various decays, see
prescriptions in Fig. \ref{Diag-fig}. 

\section{Appendix B}
Here we list the expression for the form factors of semileptonic decays $B_c
\rightarrow B_s^{(*)}$ taken from the potential model
\cite{Kis}.

\begin{eqnarray}
f_+&=&\frac{(\tilde m_c+\tilde m_s)}{2 \tilde m_s}
\sqrt{\frac{M_{B_s}}{M_{B_c}}} \xi(w), \\
f_-&=-&\frac{(\tilde m_c-\tilde m_s+2 \tilde m_b)}{2 \tilde
m_s}\sqrt{\frac{M_{B_s}}{M_{B_c}}} \xi(w), \\
F_0^A &=& \frac{M_{B_c}^2+M_{B_s^*}^2-q^2+2 M_{B_c}(\tilde m_s-\tilde m_b)}{2
\tilde m_s}
\sqrt{\frac{M_{B_s^*}}{M_{B_c}}} \xi(w), \\
F_+^A &=& -\frac{1-2\tilde m_b/M_{B_c}}{2 \tilde m_s}
\sqrt{\frac{M_{B_s^*}}{M_{B_c}}} \xi(w),
\\
F_-^A &=& \frac{1+2\tilde m_b/M_{B_c}}{2\tilde m_s}
\sqrt{\frac{M_{B_s^*}}{M_{B_c}}}
\xi(w),
\end{eqnarray}
where 
\begin{equation}
\xi(w)=\frac{2 \omega \omega_x}{\omega^2+\omega_x^2} \sqrt{\frac{2 \omega
\omega_x}{\omega^2
w^2+\omega_x^2}}\exp\left({-\frac{\tilde m_b^2(w^2-1)}{\omega^2
w^2+\omega_x^2}}\right), 
\end{equation}
\begin{equation}
\omega=2\pi \left(\frac{M_{B_c} \tilde f_{B_c}^2}{12}\right)^{1/3},\;\;\;
\omega_x=2\pi \left(\frac{M_{B_s^{(*)}} \tilde
f^2_{B_s^{(*)}}}{12}\right)^{1/3},   
\end{equation}
where $w$ is the product of $B_c$ and $B_s^{(*)}$ four-velocities. The quark
masses and the leptonic constants have the values generally used in the
calculations in the framework of potential models 
$$
\tilde m_b=4.8~\mbox{GeV},~~\tilde m_c=1.5~\mbox{GeV},~~
\tilde m_s=0.55~\mbox{GeV},~~\tilde f_{B_c}=0.47~\mbox{GeV},
~~\tilde f_{B_s^{(*)}}=0.17~\mbox{GeV}.
$$


\end{document}